\documentclass[iop, twocolumn]{aastex631}
\hypersetup{citecolor=blue,urlcolor=red}
\usepackage{multirow}
\usepackage{amsmath}
\usepackage{graphicx}
\usepackage{subfigure}
\usepackage{hyperref}  
\usepackage{nameref}
\usepackage{threeparttable}

\shorttitle{PSR~J1740+1000}
\shortauthors{Yao et al.}

\begin{document}
\title{Timing and scintillation of a young Galactic halo pulsar}
\correspondingauthor{Jumei Yao, Feifei Kou, Jianping Yuan}
\email{yaojumei@xao.ac.cn, koufeifei@xao.ac.cn, yuanjp@xao.ac.cn}

\author{J. M. Yao\href{https://orcid.org/0000-0002-4997-045X}}
\affiliation{Xinjiang Astronomical Observatory, Chinese Academy of Sciences, 150 Science 1-Street, Urumqi, Xinjiang 830011, People's Republic of China;}
\affiliation{State Key Laboratory of Radio Astronomy and Technology, Xinjiang Astronomical Observatory, CAS, 150 Science 1-Street, Urumqi, Xinjiang, 830011, P. R. China}

\author{F. F. Kou\href{**}}
\affiliation{Xinjiang Astronomical Observatory, Chinese Academy of Sciences, 150 Science 1-Street, Urumqi, Xinjiang 830011, People's Republic of China;}
\affiliation{State Key Laboratory of Radio Astronomy and Technology, Xinjiang Astronomical Observatory, CAS, 150 Science 1-Street, Urumqi, Xinjiang, 830011, P. R. China}

\author{J. P. Yuan\href{**}}
\affiliation{Xinjiang Astronomical Observatory, Chinese Academy of Sciences, 150 Science 1-Street, Urumqi, Xinjiang 830011, People's Republic of China;} 
\affiliation{State Key Laboratory of Radio Astronomy and Technology, Xinjiang Astronomical Observatory, CAS, 150 Science 1-Street, Urumqi, Xinjiang, 830011, P. R. China}

\author{Y. Wei\href{https://orcid.org/0009-0008-4753-5666}}
\affiliation{Xinjiang Astronomical Observatory, Chinese Academy of Sciences, 150 Science 1-Street, Urumqi, Xinjiang 830011, People's Republic of China;}
\affiliation{State Key Laboratory of Radio Astronomy and Technology, Xinjiang Astronomical Observatory, CAS, 150 Science 1-Street, Urumqi, Xinjiang, 830011, P. R. China}
\affiliation{School of Physical Science and Technology, Xinjiang University,Urumqi, Xinjiang, 830046, People's Republic of China}

\author{William A. Coles}
\affiliation{Electrical and Computer Engineering, University of California,\\
San Diego, 92093, USA}

\author{Richard N. Manchester}
\affiliation{Australia Telescope National Facility, CSIRO Space and Astronomy \\
P.O. Box 76, Epping NSW 1710, Australia}

\author{N. Wang}
\affiliation{Xinjiang Astronomical Observatory, Chinese Academy of Sciences, 150 Science 1-Street, Urumqi, Xinjiang 830011, People's Republic of China;}
\affiliation{State Key Laboratory of Radio Astronomy and Technology, Xinjiang Astronomical Observatory, CAS, 150 Science 1-Street, Urumqi, Xinjiang, 830011, P. R. China}

\author{S. Q. Wang}
\affiliation{Xinjiang Astronomical Observatory, Chinese Academy of Sciences, 150 Science 1-Street, Urumqi, Xinjiang 830011, People's Republic of China;}
\affiliation{State Key Laboratory of Radio Astronomy and Technology, Xinjiang Astronomical Observatory, CAS, 150 Science 1-Street, Urumqi, Xinjiang, 830011, P. R. China}

\author{W. M. Yan}
\affiliation{Xinjiang Astronomical Observatory, Chinese Academy of Sciences, 150 Science 1-Street, Urumqi, Xinjiang 830011, People's Republic of China;}
\affiliation{State Key Laboratory of Radio Astronomy and Technology, Xinjiang Astronomical Observatory, CAS, 150 Science 1-Street, Urumqi, Xinjiang, 830011, P. R. China}

\begin{abstract}
We conducted a timing and scintillation study of a young Galactic halo pulsar, PSR~J1740+1000, using observations from the Nanshan, FAST, and Parkes radio telescopes. From a timing analysis, we measured the proper motion of this pulsar for the first time. The proper motion measurement indicates that the pulsar is moving away from the Galactic plane at an position angle of 16$^\circ.7\pm4^\circ.8$ (in Galactic coordinates), with a total proper motion of 56.9$\pm$8.0~mas/yr and a corresponding transverse velocity of 329$\pm$80~km/s. This velocity suggests that PSR~J1740+1000 is a typical-velocity, young pulsar born within the Galactic halo. For scintillation, we detected scintillation arcs, arclets and double-layered adjacent arcs in the secondary spectra. Under the isotropic and anisotropic scattering cases, the screen-to-pulsar distance is 370$\pm$72 and 1$\pm$12~pc, respectively. The latter value closely matches the scale of the Pulsar Wind Nebula (PWN) associated with PSR~J1740+1000 and provides a better fit, suggesting that its scattering is most likely dominated by the PWN. The double-layered adjacent arcs observed on MJD~60180 imply that the pulsar’s scattered image consists of two dominant components (A and B), along with multiple weaker components. Component A is located at the pulsar's geometric position with an angular position of 0~$\mu$as, while Component B is located 112$\pm$16 and 23$\pm$17~$\mu$as away from the central component under the isotropic and anisotropic scattering cases, respectively. As the angular position of Component B is independent of observing frequency, this hints at refraction by an AU-scale structure located within the scattering region, which may originate from the PWN.
\end{abstract}
 
\section{Introduction}\label{sec:Intro}
Pulsars are born with natal kick velocities resulting from asymmetric supernovae explosions. According to the pulsar velocity distribution given by \cite{hll+05}, birth velocities typically range from about 100 to 500~km/s, with a small fraction exceeding 1000~km/s, known as hyper-velocities. Regardless of the exact form of the pulsar velocity distribution, it is widely accepted that a significant number of pulsars can escape the Galactic gravitational potential and may be found far beyond the Galactic disc \citep{bai19, nsk+13, rcl+18}. Unlike older halo pulsars, the existence of young pulsars in the Galactic halo can be attributed to a hyper-velocity pulsar, i.e. greater than 1000~km/s, escaping from the Galactic disk or their formation within the halo itself. The velocities of young halo pulsars generated by these two mechanisms differ significantly in both magnitude and direction. In terms of direction, young halo pulsars produced by the first mechanism have velocities pointing away from the Galactic plane, whereas those from the second mechanism have an equal probability of moving toward or away from the plane. Regarding magnitude, young halo pulsars from the first mechanism have hypervelocities, while those from the second mechanism do not necessarily exhibit hypervelocity. Therefore, measuring the velocities of halo pulsars is essential for understanding their origin.  

PSR~J1740+1000, discovered in an Arecibo survey at 430~MHz \citep{mac+02}, is an example of a young pulsar situated at a high Galactic latitude. From follow-up timing observations conducted at Arecibo, \cite{mac+02} determined that the spin-down age, or the characteristic age, of PSR~J1740+1000 is approximately 114~kyr, which is less than that of 95\% of all known pulsars. Its distance of 1.22$\pm$0.24~kpc, estimated from the dispersion measure using the YMW16 Galactic electron-density model \citep{ymw16} and assuming 20\% uncertainties, along with its high Galactic latitude (Gb$=$20.3$\degr$), suggests that this young pulsar is located 418~pc above the Galactic plane. If we assume that the pulsar was born in the midplane of the Galaxy and that its kinematic age is equal to its spin-down age, then PSR~J1740+1000 would be classified as a hyper-velocity pulsar with a velocity greater than 3500~km/s. As mentioned, young pulsars situated high above the Galactic plane, such as PSR~J1740+1000, suggest either a progenitor from the halo population or a neutron star (NS) that was ejected from the Galactic disk with a hyper-velocity. Where did PSR~J1740+1000 come from? Based on Arecibo observations conducted at seven epochs centered at 1.4~GHz, \cite{mac+02} performed an autocorrelation function analysis of the detected dynamic spectra. They obtained the scintillation timescale and bandwidth, finding that the mean scintillation velocity is approximately 250~km/s. They also suggested that the scattering may be dominated by the North Polar Spur (NPS) or the Gould Belt, as the line of sight to PSR~J1740+1000 passes through these structures, both of which are located within 300~pc of Earth. Assuming the scintillation velocity is the pulsar velocity, they predicted that PSR~J1740+1000 was formed through accretion induced collapse (AIC) of a halo white dwarf. Such assumptions come with significant uncertainties. As mentioned in \cite{mac+02}, the strong refractive scintillation influences the measurements of scintillation parameters and scintillation velocity. Additionally, the unknown position of the scattering screen introduces further uncertainty in the conversion between scintillation velocity and pulsar velocity. 

Using X-ray observations from Chandra and XMM-Newton, \cite{kmp+08} studied the pulsar wind nebulae (PWN) of PSR~J1740+1000, with a particular focus on the extended long tail. From the orientation of this tail, shown in their Figure~3, \cite{kmp+08} determined that the pulsar is moving at a very small angle of about 7$\degr$ toward the Galactic plane, suggesting that PSR~J1740+1000 originated from a halo-star progenitor, born out of the Galactic plane. Assuming that the observed tail is a result of the pulsar's motion, the position angle ($\Psi_{\rm pm}$) of the pulsar's proper motion is about 35$\degr$ (i.e., the angle in the sky plane, measured from north toward east). Combining the centroid position detected by Chandra with a 10-year separation and the nominal Chandra position uncertainty, \cite{hbg13} placed an upper limit on both the proper motion of 60~mas/yr and the transverse velocity of 340d$_{1.2}$~km/s, where  d$_{1.2}$ is the distance in units of 1.2~kpc. There are only a few young halo pulsars similar to PSR~J1740+1000. Examples include PSR~B0919+06 \citep{ccl+01}, 1RXS J141256.0+792204, also known as Calvera \citep{mrt+21} and PSR~J0837$-$2454 \citep{Pol_2021}.  

Pulsar proper motions can be measured using various methods, including Very Long Baseline Interferometry (VLBI), pulsar timing, scintillation, and other techniques. Among these, VLBI and pulsar timing are the most reliable. For pulsars, timing analysis can be used to determine the spin period, period derivative, sky position, distance, proper motion, and, for young pulsars, glitch events \citep{dym+20}. When the timing data span many years and have good cadence, the proper motion can be obtained through a global timing fit \citep{jsk+08, jsb+10}. For young pulsars, however, discontinuous datasets or the presence of glitches often limit the effectiveness of global fitting. In such situations, the proper motion can instead be derived by fitting the positions measured at multiple epochs as a function of time \citep{skz+19, wwy+25}. Based on long-term observations of 15 pulsars, \cite{lwy+16} found that proper motions obtained from position–time fitting are generally consistent with those derived from global timing fits. More recently, \cite{wwy+25} showed that proper motions derived from FAST and Fermi-LAT multi-epoch positions agree with VLBI measurements within 0.24$\sigma$ \citep{bst+23}. Among the parameters obtained from pulsar timing, the distance and proper motion are essential for determining the location of the scattering screen in pulsar scintillation studies. Glitches are abrupt increases in a pulsar's rotation rate, occurring as rare and irregular events. According to the glitch table \footnote{http://www.jb.man.ac.uk/pulsar/glitches.html} compiled by \cite{bsa+22}, their magnitudes, characterized by $\Delta\nu/\nu$, span a wide range from approximately $10^{-10}$ to $10^{-5}$. Glitches are believed to result from either the sudden transfer of angular momentum from the interior superfluid to the crust of the pulsar \citep{cha13} or from crustquakes within neutron stars \citep{bbl20}. For PSR~J1740+1000, based on observations from 76-m Lovell Telescope at Jodrell Bank two glitches were detected at MJDs~54747.6 \citep{els+11} and 56164 \citep{bsa+22}, with magnitudes of $10^{-9}$ and $10^{-6}$, respectively.

Pulsar interstellar scintillation (ISS), which includes diffractive scintillation (DISS) and refractive scintillation (RISS), is a powerful probe for studying both the ionized interstellar medium (IISM) and pulsars. DISS, associated with small-scale turbulence in the IISM, appears in dynamic spectrum as intensity fluctuations known as scintles and in secondary spectrum as scintillation arcs \citep{smc+01}. In contrast, RISS, caused by large-scale turbulence, modulates DISS and manifests in dynamic spectrum as periodic fringes or criss-cross patterns, and in secondary spectrum as hot spots or reverse arclets \citep{cw86, hsa+05}. Currently, reverse arclets have been observed in several pulsars. In the simulations of \cite{crs+06}, it was shown that for anisotropic scattering with an axial ratio of 4:1, where the scattering spectrum is elongated along the velocity direction, the resulting arc structure consists of finely spaced parallel arcs, or multiple adjacent arc layers, that are further intersected by reverse arclets (see their Figure 11). As discussed in \cite{crs+06}, two mechanisms have been proposed to explain the origin of these reverse arclets and multiple adjacent arc layers: one involves stochastic, speckle-like substructures in the scattered image produced by a turbulent screen with a Kolmogorov spectrum (although not all pulsars exhibit a Kolmogorov scintillation spectrum; see \cite{bts+19}), and the other involves discrete features in the scattered image caused by refraction from compact, dense AU-scale structures. These mechanisms result in different frequency dependencies of the angular positions of these sub-images ($\theta$): in the first case, $\theta \propto \nu^{-2}$, while in the second, the small size of the structure leads to a frequency-independent $\theta$. Therefore, measuring the frequency dependence of $\theta$ can help distinguish between the two mechanisms. In high–signal-to-noise observations of certain pulsars that exhibit reverse arclets, both the reverse arclets and multiple adjacent arc layers can be seen simultaneously in the secondary spectra, such as in PSR~B1133+16 \citep{sro+19} and PSR~B1737+13 \citep{csb+25}. Owing to the significant influence of the reverse arclets, these studies did not undertake separate curvature measurements for the individual adjacent arc layers, nor did they investigate the frequency evolution of their associated $\theta$ values. Estimating the curvature of reverse arclets is more challenging than for arcs. To address this issue, \cite{swm+21} and \cite{bbk+22} developed a method for measuring the curvature of reverse arclets from the $\theta$-$\theta$ space. Using FAST, \cite{occ+24} conducted a single observation of PSR~J1740+1000 at MJD~59510 and detected the scintillation arc and arclets for the first time. Since the pulsar's transverse velocity is unknown, a single measurement of the arc curvature is insufficient to constrain the location of the scattering screen. For a given arc curvature, it could be explained by a combination of a large transverse velocity and a small Earth-to-screen distance, or vice versa. Therefore, with the proper motion known, the arc curvature from multiple observations covering Earth's orbit can be used to determine the location and uncover the turbulent properties of the scattering screen.     
 
In this paper, we use observations from Nanshan, Parkes Murriyang and FAST to measure the proper motion and study the scintillation of PSR~J1740+1000. The paper is structured as follows: in Section~\ref{sec:Obs}, we describe the Nanshan, Parkes and FAST observations and data processing methods; in Section~\ref{sec:TOA}, we present the timing results, including the detection of a glitch and measurement of the proper motion; in Section~\ref{sec:ISS}, we present the observed dynamic spectra along with the autocorrelation function analysis, measure the arc curvature, analyze the properties of the scattering screen, and analyze the double-layered adjacent arc detected in the secondary spectra; finally, in Section~\ref{sec:SC}, we summarize our findings and draw conclusions.

\section{Observations and data processing}\label{sec:Obs}
Pulsar timing observations at the Xinjiang Astronomical Observatory, Chinese Academy of Sciences, began in January 2000 using the 25-m telescope at the Nanhsan station. This telescope underwent a comprehensive upgrade between 2014 and 2018, and a new 26-m telescope became operational in 2019. A cryogenic receiver equipped with an analogue filterbank (AFB) was employed to obtain data before 2010 \citep{wmz+01}. Beginning in January 2010 , a digital filterbank (DFB) came into operation. Since 2020, a pulsar digital-backend developed on ROACH2 (Re-configurable Open Architecture Computing Hardware-2) platform was put into operation \citep{haz+16}. A wider L-band receiver was commissioned in 2023, together with a new digital backend based on the RFSoC platform \citep{lcd+25}. The data spans and the corresponding backend and receiver configurations for the Nanshan telescope (abbreviated as NS) are listed in Table~\ref{tab:backend}. PSR~J1740+1000 was typically observed with two sessions per month, with integration times of 16~min per session. As mentioned in Section~\ref{sec:ISS}, the observed flux of PSR~J1740+1000 varies from day to day, resulting in only about one-third of the data being suitable for timing study. In this paper, we present two segments of timing observational data for this pulsar from Nanshan: one spanning from 2008 to 2013 (MJD 54620–56645) and the other from 2020 to 2024 (MJD 58927–60412). We processed the data from Nanshan, FAST, and Parkes using the {\sc dspsr} analysis program \citep{sb11}\footnote{\url{http://dspsr.sourceforge.net}} and the {\sc psrchive} software package \citep{sdo12}\footnote{\url{http://psrchive.sourceforge.net}}. The data from Nanshan were integrated off-line over time, frequency, and polarization to generate total intensity profiles after removing radio frequency interference (RFI).

Using FAST, we observed PSR~J1740+1000 across twelve epochs, covering different phases of Earth's orbit. The Modified Julian Date (MJD), day of year, and corresponding observation duration for each epoch are listed in columns 1, 2, and 3 of Table~\ref{tab:arc_curvature}, respectively. Since the observations at MJDs~59945, 60308, and 60309 correspond to nearly the same orbital phase, and the observations at MJDs 60036 and 60037 also correspond to nearly the same phase, these observations effectively cover nine distinct phases of Earth's orbit. As shown in Table~\ref{tab:backend}, for FAST, the 19-beam receiver operates over 1000–1500~MHz. Owing to reduced sensitivity near the band edges, the effective usable bandwidth is 400~MHz, covering from 1050 to 1450~MHz. The data for each channel, with a bandwidth of 0.122~MHz, were folded at the topocentric pulse period using a sub-integration time of 10~s. After removing RFI, the data were polarization calibrated.

From the Parkes archive data \footnote{\url{https://data.csiro.au/domain/atnf}}, there are six approximately 1-hour observations of PSR~J1740+1000 conducted on MJDs 57663, 57731, 57776, 57814, 57837 and 57841. These observations were recorded using the PDFB4 receiver. Except for the observation on MJD~57663, the data for the remaining observations was recorded with a bandwidth of 256~MHz, centered at 1369~MHz, and divided into 2048 channels, resulting in a channel bandwidth of 0.125~MHz. For the observation on MJD~57663, the data was recorded with a bandwidth of 512~MHz, centered at 1465~MHz, using 2048 channels. We also listed data spans and the corresponding backend and receiver configurations for the Parkes telescope in Table~\ref{tab:backend}. As discussed in Section~\ref{sec:ISS}, the observed flux of PSR~J1740+1000 varies from day to day, and only four of these observations, conducted on MJDs~57731, 57776, 57814 and 57841, were suitable for the scintillation study. The data for these observations were already folded at the topocentric pulse period with a sub-integration time of 20~s. After downloading the data, we were only able to perform polarization calibration for five of the observations on MJDs~57663, 57776, 57814, 57837, and 57841, and flux calibration for the last three of these observations.

\begin{table*}
\caption{Data spans and corresponding backend and receiver configurations for all Nanshan, FAST, and Parkes observations analyzed in this study.}
\centering
\setlength\tabcolsep{6.0pt}
\footnotesize
\begin{tabular}{cccccccc}
	\hline
	data span (MJD)& data span& backend & central frequency & bandwidth & Nchan\\ 
	\hline
	&  & & MHz& MHz& \\
	\hline
    54620-55196 (NS)& 2008.06-2009.12 & AFB & 1540 & 320 & 128 \\
    55197-56645 (NS)& 2010.01-2013.12 & DFB & 1540 & 512 & 1024 \\
    58927-59755 (NS)& 2020.03-2022.06 & ROACH2 & 1540 & 512 & 512\\
    60043-60413 (NS)& 2023.04-2024.04 & RFSoc & 1412 & 600 & 600 \\
    59510-60494 (FAST)&2021.10-2024.07 &  ROACH2& 1250 & 500 & 4096 \\
    57663-57841 (Parkes) &2026.10-2017.03 & PDFB4 & 1369, 1465& 256, 512&2048\\
	\hline
	\end{tabular}
	\label{tab:backend}
\end{table*}

For the timing study, we used data from both Nanshan and FAST. The total intensity profiles were then cross-correlated with standard profiles to obtain local pulse times-of-arrival (ToAs). Finally, local ToAs were corrected to the Solar system using the standard timing program TEMPO2 \citep{hem06} with the Jet Propulsion Laboratories planetary ephemeris DE~421 \citep{fwb+09}. For the Nanshan data, as well as for the joint analysis of Nanshan and FAST observations, the TOAs from different backends exhibit systematic offsets. Therefore, in our data-processing pipeline, we generate standard profiles and corresponding TOAs separately for each backend, and then apply relative offsets to remove the systematic differences. The ToAs were weighted by the inverse square of their uncertainties. The pulse phase $\Phi$ of the of the standard timing model is:
\begin{equation}
\Phi(t)=\Phi_{0}+\nu(t-t_0)+\frac{1}{2}\dot{\nu}(t-t_0)^2+\frac{1}{6}\ddot{\nu}(t-t_0)^3+...
\end{equation}
where $\Phi_{0}$ is the the phase at time $t_0$, $\nu$, $\dot{\nu}$ and $\ddot{\nu}$ are the phase frequancy and its derivatives, respectively.

Pulsar glitch is characterized by advance in pulse arrival times, and the corresponding phase variation can be expressed as:
\begin{equation}
\phi_{g}=\Delta{\phi}+\Delta{\nu}_{p}(t-t_{g})+\frac{1}{2}\Delta \dot{\nu}_{p}(t-t_{g})^{2}+[1-e^{(t_{g}-t)/\tau_{d}}]\Delta{\nu}_{d}\tau_{d} \ ,
\label{equ:glitchmodel}
\end{equation}
where $\Delta{\phi}$ is the phase difference between pre- and post-glitch, $t_{g}$ is the epoch of the glitch, $\Delta{\nu}_{p}$ and $\Delta{\dot{\nu}}_{p}$ are the permanent changes in rotation frequency and its derivative caused by glitch, $\Delta\nu_d$ is the amplitude of the
exponential decay on a time-scale $\tau_d$.

For the scintillation study, we used data from both FAST and Parkes. Following the data processing procedures outlined in \cite{yzm+21}, we obtained the 2D dynamic spectra (power as a function of radio frequency and time) for each observation. For the FAST data, because the two consecutive segments of the observational data, from 1050 to 1150~MHz and from 1300 to 1450~MHz, were not affected by RFI, we selected these two segments for the subsequent auto-correlation function and secondary spectrum analysis. For the Parkes data, we excluded the outer 25~MHz of the frequency band and used data from 1266 to 1472~MHz for the dynamic spectrum and secondary spectrum analysis. By performing Fourier Transform analysis on the dynamic spectrum, we obtained the secondary spectrum. The axes of the secondary spectrum are conjugate time ($f_t$, in mHz), equivalent to a differential Doppler shift, and conjugate frequency ($f_\nu$, in $\mu$s), equivalent to a differential delay. For FAST data, the Nyquist frequencies for the channel bandwidth and sub-integration time are $f_t$(Nyquist)$=$50~mHz and $f_\nu$(Nyquist)$=$4.096~$\mu$s. For Parkes data, the Nyquist frequencies are $f_t$(Nyquist)$=$25~mHz and $f_\nu$(Nyquist)$=$4.0~$\mu$s.

\section{The timing of PSR~J1740+1000}\label{sec:TOA}
By combining data from Nanshan and FAST, we detected a significant glitch and determined the proper motion of PSR~J1740+1000. In Appendix~\ref{sec:glitch}, we describe the glitch in the Nanshan data around MJD~56167, with a magnitude of 2899.2(8)$\times$10$^{-9}$ and a recovery timescale of 93(17) days, which closely matches the large glitch detected in the Jodrell Bank data \citep{bsa+22}. In the following section, we present our analysis results for the proper motion.

\subsection{The proper motion of PSR~J1740+1000}
To determine the proper motion of PSR~J1740+1000, following \cite{wwy+25}, we performed timing analysis on observations from Nanshan and FAST using both the classic pulsar timing software package TEMPO2 and the modern pulsar timing software package PINT \citep{lrd+21}. According to \cite{lrd+21}, TEMPO2 and PINT have been tested and are generally shown to produce nearly identical results. Therefore, employing both packages facilitates cross‑checking the timing analysis. As shown in Table~\ref{tab:position}, \cite{mac+02} previously reported a position at MJD~51662, derived from approximately two years of Arecibo observations, spanning from May 1999 to April 2001, through timing analysis using TEMPO. Since Nanshan has conducted nearly continuous observations of PSR~J1740+1000 from 2008 to 2012 and from 2020 to 2024, we divided the Nanshan data into these two segments for processing. During the timing analysis of each data segment, to determine the position of PSR~J1740+1000, we fit higher-order spin parameters to mitigate effects of timing noise. This approach has been widely adopted as an empirical method to whiten timing residuals in pulsars affected by strong timing noise (e.g., \cite{zhw+05}). To assess the impact of timing noise and the inclusion of higher-order spin-frequency derivatives on position fitting, we carried out a dedicated test using PSR~J1906+0746, which has a period and period derivative comparable to those of PSR~J1740+1000. This test demonstrates that the fitted position converges once sufficiently high-order derivatives are included (see Appendix~\ref{sec:fit_simulation} for details). For PSR~J1740+1000, we adopted the same strategy to determine the pulsar position by fitting higher-order spin-frequency derivatives. The criterion for selecting a sufficiently high order is that the fitted position becomes stable and the uncertainties of the highest-order spin-frequency derivatives remain smaller than their fitted values. After excluding the data affected by the large glitch in the first data segment, we determined the position at MJD~55384 by fitting up to the fourth spin-frequency derivative (F4) using both TEMPO2 and PINT. The corresponding fitting results are shown in the upper part of Table~\ref{tab:p_result} in Appendix~\ref{sec:fit_simulation} and are also listed in the second and third rows of Table~\ref{tab:position}. By combining FAST observations with the second segment of the Nanshan data, we obtained the position at MJD~59711 by fitting up to the third spin-frequency derivative (F3) using both TEMPO2 and PINT. The fitting results are presented in the lower part of Table~\ref{tab:p_result} in Appendix~\ref{sec:fit_simulation} and summarized in the last two rows of Table~\ref{tab:position}. For the same set of data, the position results obtained by PINT and TEMPO2 are completely identical.

\begin{figure}[!h]
\center
 \includegraphics[width=8.0 cm, angle=270]{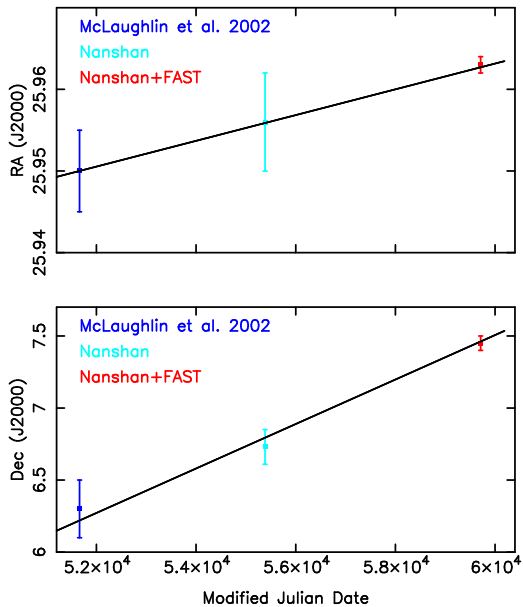}
    \caption{The position of PSR~J1740+1000 at MJDs~51662, 55384 and 59711. The blue points represent positions derived from AO observations, the cyan points represent positions obtained from the first segment of Nanshan data, and the red points represent positions derived from the second segment of Nanshan data combined with FAST observations. The black lines indicate the best-fit results.} 
\label{fig:pm}
\end{figure}

The three positions are shown in Figure~\ref{fig:pm}. To estimate the proper motion in Right Ascension, $\mu_\alpha$, and Declination, $\mu_\delta$, we use the curve-fit function from Python's Scipy library \footnote{\url{https://docs.scipy.org/doc/scipy/index.html}} to perform a linear weighted least-squares fit. The fitting results in 
\begin{equation}
\mu_\alpha=8.5\pm3.2~\rm mas/yr
\label{eq:PMRa}
\end{equation}
and 
\begin{equation}
\mu_\delta=56.3\pm8.0~\rm mas/yr.
\label{eq:PMDec}
\end{equation}
The total proper motion, $\mu_{\rm tot}$, is 56.9$\pm$8.0~mas/yr. In Figure~\ref{fig:pm}, the black line represents the best-fit results. Based on the measured proper motion, the position angle, $\Psi_{\rm pm}$, is 8$^\circ.6\pm1^\circ.2$. In Galactic coordinates, the proper motion of PSR~J1740+1000 points 16$^\circ.7\pm4^\circ.8$ away from the Galactic disk.   

\begin{table*}
\caption{The position of PSR~J1740+1000 in J2000 Equatorial Coordinates measured through timing.}
\centering
\setlength\tabcolsep{2.0pt}
\footnotesize
\begin{tabular}{cccccccc}
	\hline
	Right Ascension& Declination & POSEPOCH & Data Range &No. of TOAs & Postfit rms timing residual& Telescope & Software\\ 
	\hline
	h:m:s & d:m:s & MJD & MJD &  & $\mu s$ &  &  \\
	\hline
   17:40:25.9500(50)&+10:00:06.30(20)&51662&51310-52014&684&952&Arecibo&TEMPO\\
   17:40:25.9559(67)&+10:00:06.74(12)&55384&54620-56148&72&965&NS&TEMPO2\\
   17:40:25.9559(67)&+10:00:06.74(12)&55384&54620-56148&72&965&NS&PINT\\
   17:40:25.9627(17)&+10:00:07.47(7)&59711&58927-60494&43&197&NS+FAST&TEMPO2\\
   17:40:25.9627(17)&+10:00:07.47(7)&59711&58927-60494&43&197&NS+FAST&PINT\\
	\hline
	\end{tabular}
	\label{tab:position}
\end{table*}

\section{The scintillation of PSR~J1740+1000}\label{sec:ISS}
Using data from FAST and Parkes, we observed both periodic fringes and criss-cross patterns in the dynamic spectra and detected scintillation arcs, arclets, and double-layered adjacent arcs in the secondary spectra for PSR~J1740+1000. In this section, we present the dynamic spectra, provide measurements of the arclet curvatures, derive the location of the scattering screen, and perform an analysis of the double-layered adjacent arcs.

\subsection{The dynamic spectra}\label{sec:DS}
\begin{figure*}[!ht]
\center
 \includegraphics[width=14.0 cm, angle=0]{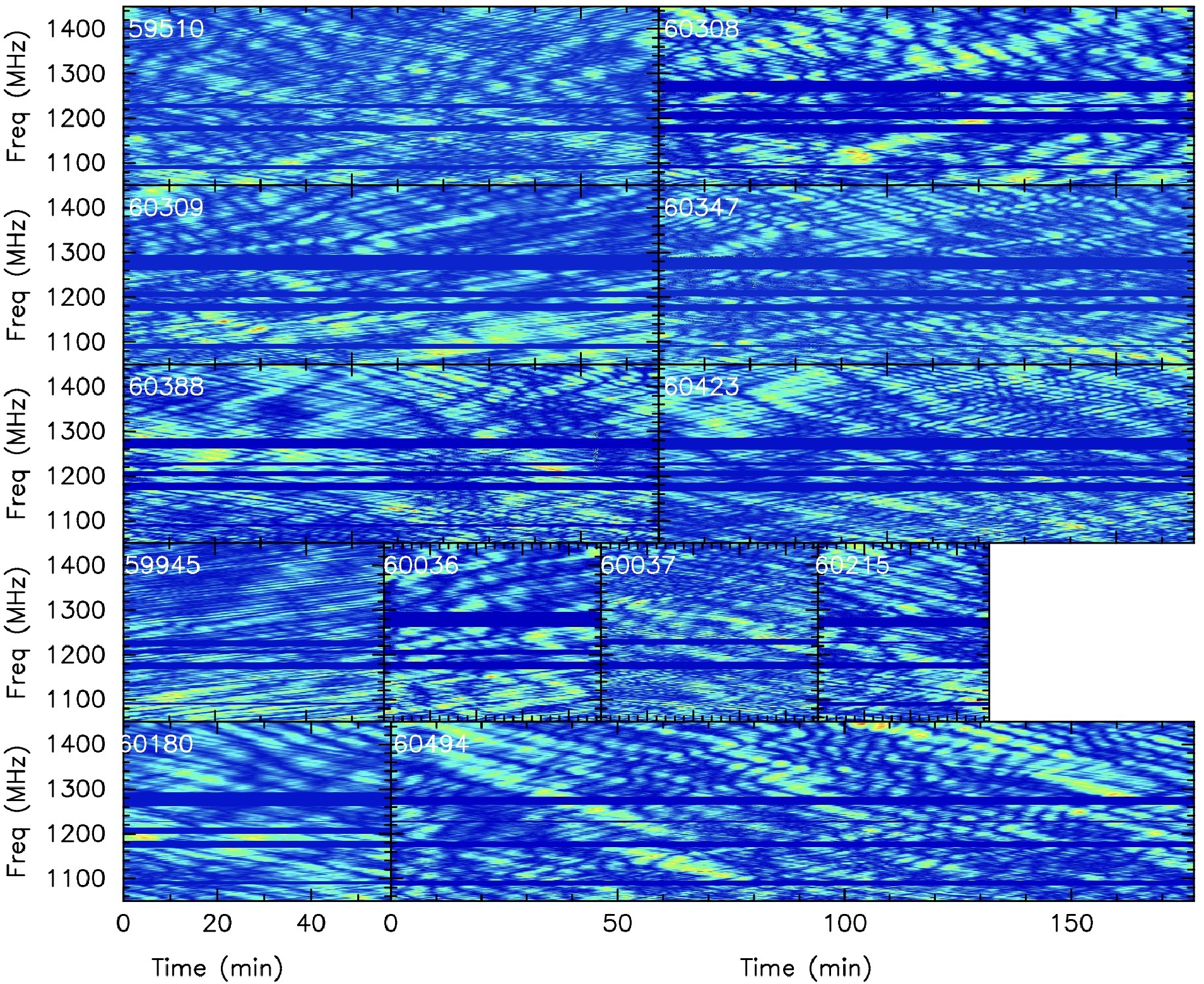}
    \caption{The dynamic spectra from twelve observations of PSR~J1740+1000 using FAST centered at 1250~MHz.}
    \label{fig:FDS}
\end{figure*}

\begin{figure*}[!ht]
\center
 \includegraphics[width=10.0 cm, angle=270]{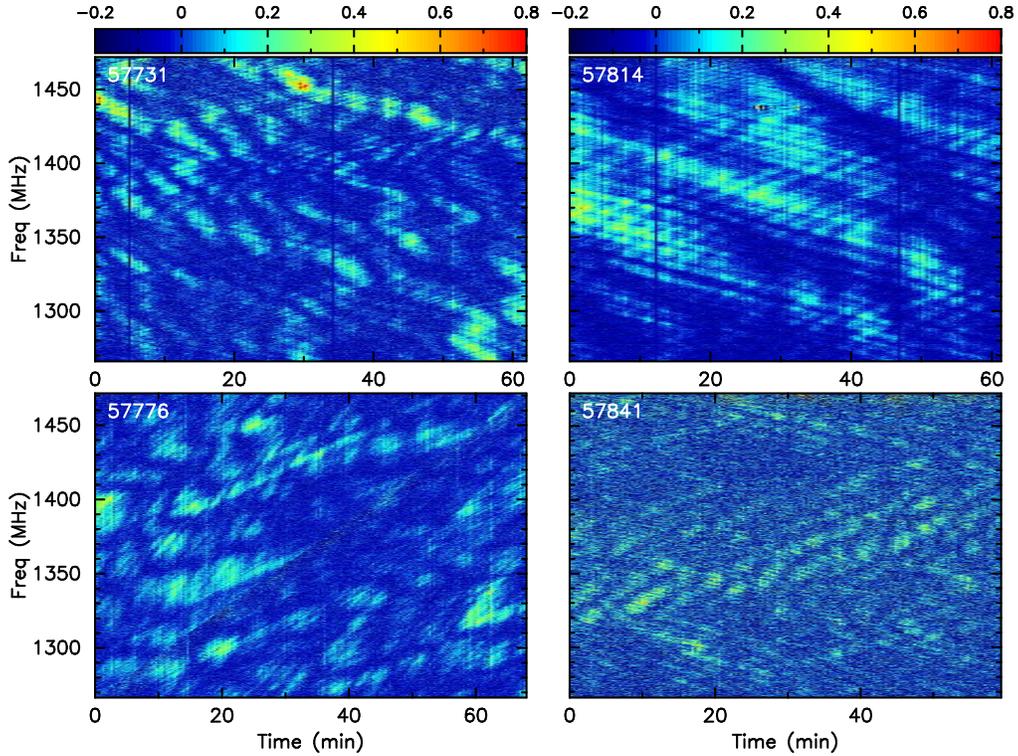}
    \caption{The dynamic spectra from four observations of PSR~J1740+1000 using Parkes centered at 1369~MHz.}
    \label{fig:PDS}
\end{figure*}

\begin{figure*}[!ht]
\center
 \includegraphics[width=8.0 cm, angle=270]{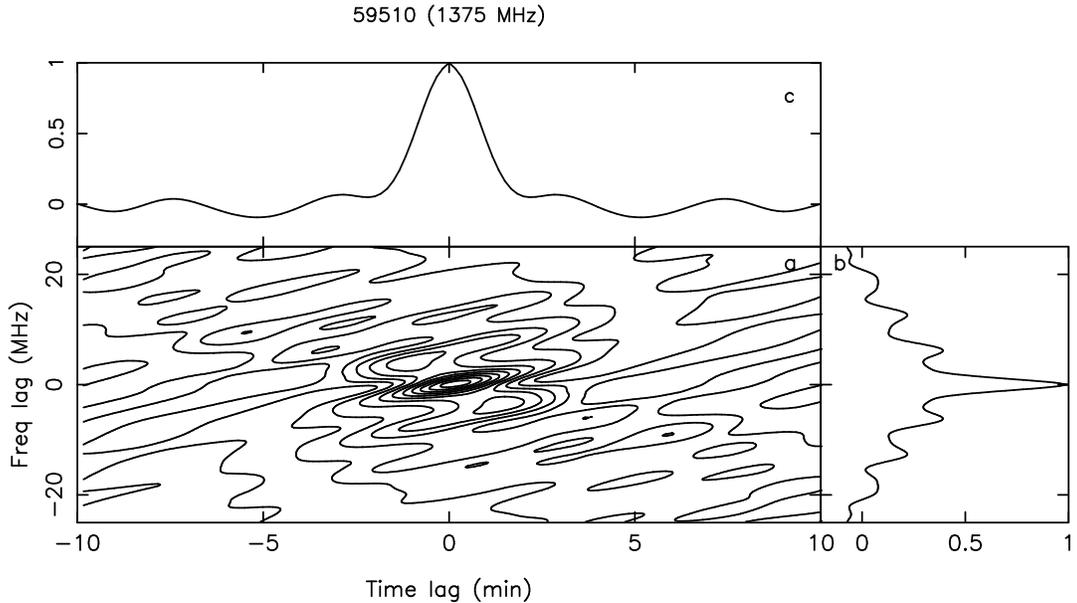}
    \caption{The ACF of PSR~J1740+1000 on MJD~59510 at 1375~MHz. Panel a: the 2D ACF. Panel b: the corresponding 1D frequency-domain ACF. Panel c: the corresponding 1D time-domain ACF.}
    \label{fig:acf}
\end{figure*}

We present the dynamic spectra of PSR~J1740+1000 from twelve FAST observations in Figure~\ref{fig:FDS} and from four Parkes observations in Figure~\ref{fig:PDS}. The dynamic spectra from FAST (1050-1450~MHz) and Parkes (1266-1472~MHz) display a periodic fringe or criss-cross pattern. Unlike the coherent superposition of a single image, where the dynamic spectrum can be characterized by the scintillation bandwidth ($\Delta \nu_d$) and timescale ($\Delta t_d$), the periodic fringe and criss-cross pattern arises from interference between each discrete sub-image and the rest of the image. As a result, we cannot accurately determine the scintillation bandwidth and timescale through autocorrelation function (ACF) fitting. If the pulsar scattered image is dominated by two main sub-images, we can detect a noticeable periodic modulation in the ACF. As shown in Figure~\ref{fig:acf}, both the frequency-domain and time-domain 1D ACF exhibit clear periodic modulation, with the modulation in the frequency domain being more pronounced than in the time domain. The origin of this phenomenon will be discussed in Section~\ref{sec:double_arcs}

For PSR~J1740+1000, observations from Parkes, Nanshan, and FAST show significant intensity modulation. For Parkes observations, we were able to obtain flux measurements for only three epochs: MJDs~57814, 57837 and 57841. After flux calibration, the flux values were found to be approximately 11.3, 1.7, and 2.7~mJy, respectively. Thus, the observed flux varied by a factor of about 7. FAST observations include two sets of two consecutive days of data; however, since FAST is currently unable to perform flux calibration, we can only estimate the relative intensity using the radiometer equation from \cite{lk04}. Our analysis shows that the relative intensity varied from 1.87(4) to 7.46(8). Additionally, for the data collected on consecutive days, the maximum intensity varies by approximately a factor of 1.5. Since PSR~J1740+1000 doesn't exhibit obvious mode changes and its intensity varies from day to day, it is highly likely that this variation is caused by RISS. Since we are unable to obtain scintillation parameters from the ACF, we cannot derive the timescale of the RISS. Future continuous observations over multiple days, which provide intensity modulation timescales, may help identify the cause of the intensity variations.

\subsection{The secondary spectra and measurement of the arc curvature}\label{sec:ss_arc}
\begin{figure*}[!ht]
\center
 \includegraphics[width=9.5 cm, angle=270]{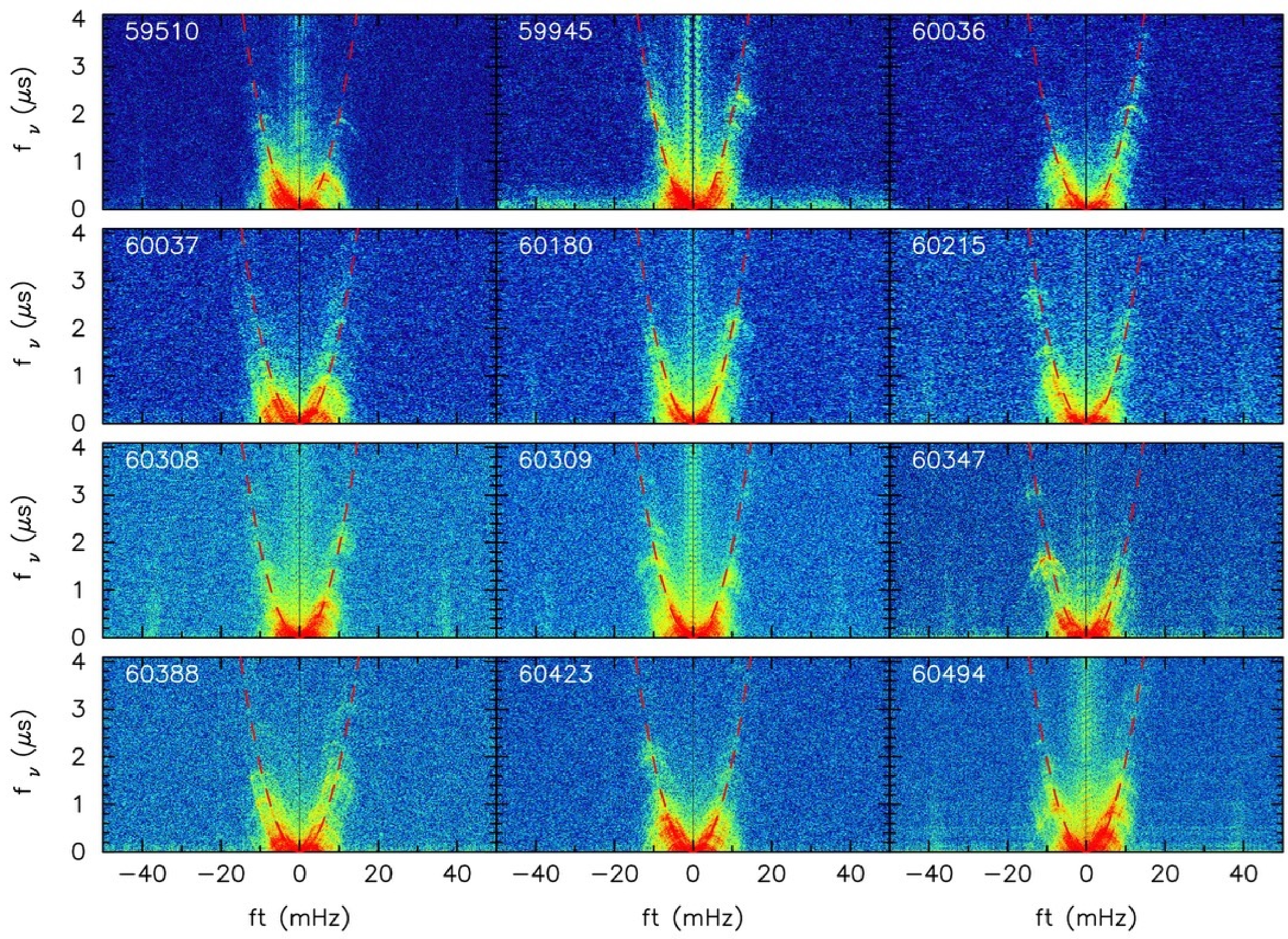}
    \caption{Secondary spectra for PSR~J1740+1000 from twelve FAST observations. The red dash lines represent the best-fit arc curvature cetered at 1100~MHz.}
\label{fig:ss_1100}
\end{figure*}

\begin{figure*}[!ht]
\center
 \includegraphics[width=9.5 cm, angle=270]{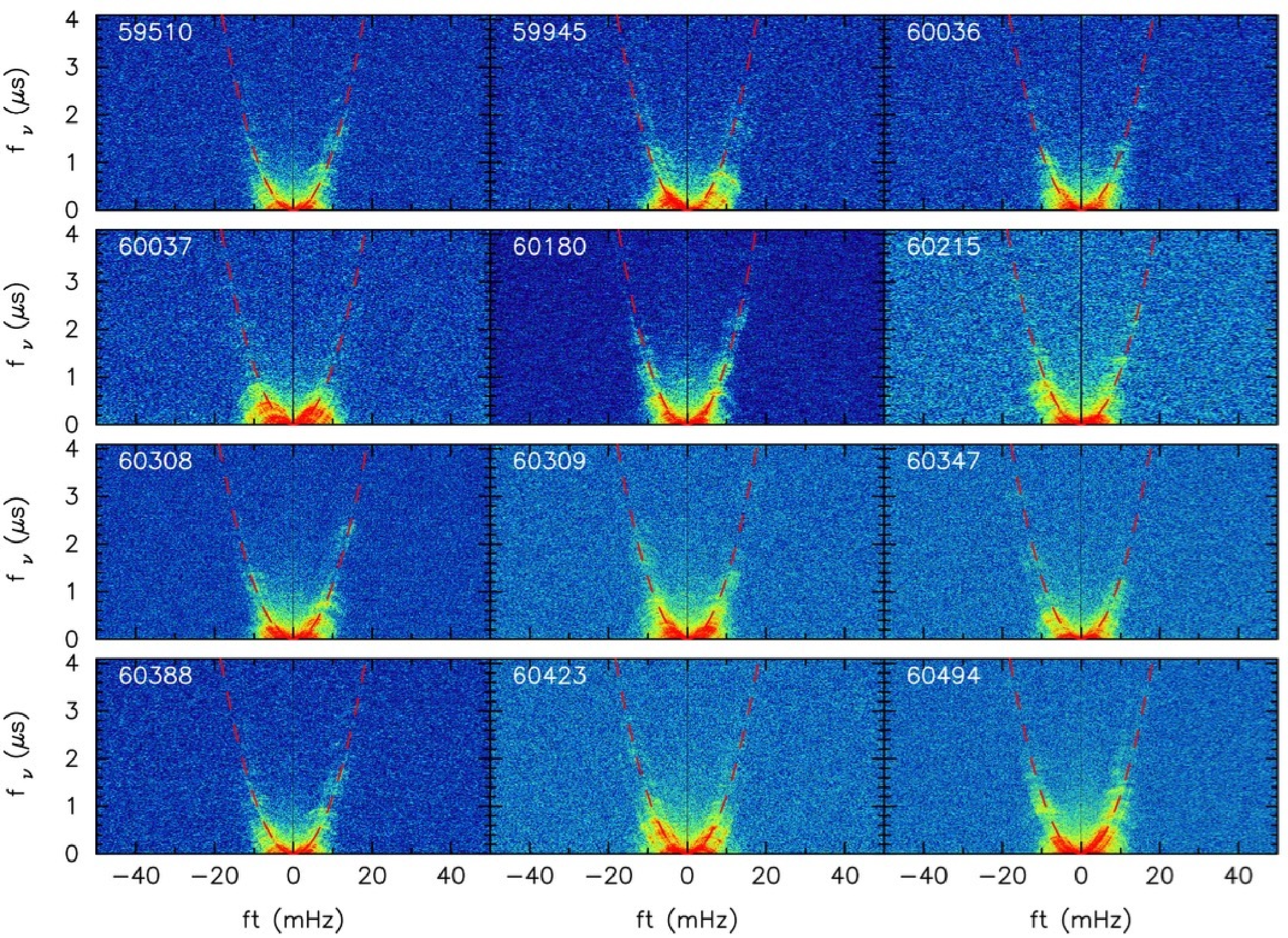}
    \caption{Secondary spectra for PSR~J1740+1000 from twelve FAST observations. The red dash lines represent the best-fit arc curvature cetered at 1375~MHz.}
 \label{fig:ss_1375}
\end{figure*}

\begin{figure}[!ht]
\center
 \includegraphics[width=6.0 cm, angle=270]{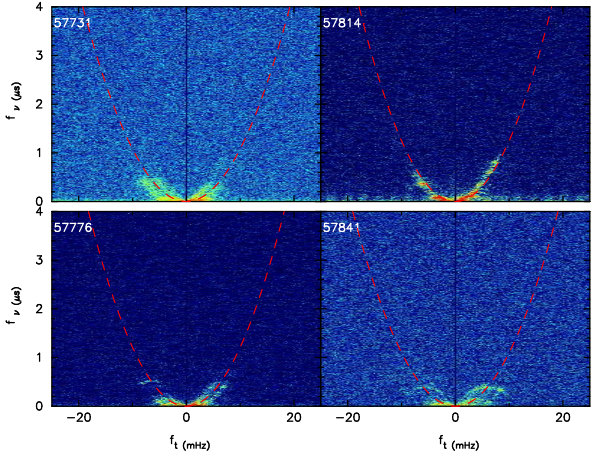}
    \caption{Secondary spectra for PSR~J1740+1000 from four Parkes observations. The red dash lines show the best-fit arc curvature cetered at 1369~MHz.}
\label{fig:ss_1369}
\end{figure}

In Figures~\ref{fig:ss_1100}, \ref{fig:ss_1375} and \ref{fig:ss_1369}, we show the secondary spectra of PSR~J1740+1000 centered at 1100, 1375 and 1369~MHz from twelve FAST observations and four Parkes observations. The secondary spectra from all these epochs consistently show reverse arclets. And for several epochs, the detected primary arc consists of double-layered adjacent arcs, such as the arc detected on epoch MJD~60180 at 1100~MHz. A detailed discussion of this is provided in Section~\ref{sec:double_arcs}. 

\begin{figure}[!ht]
\center
 \includegraphics[width=8.0 cm, angle=0]{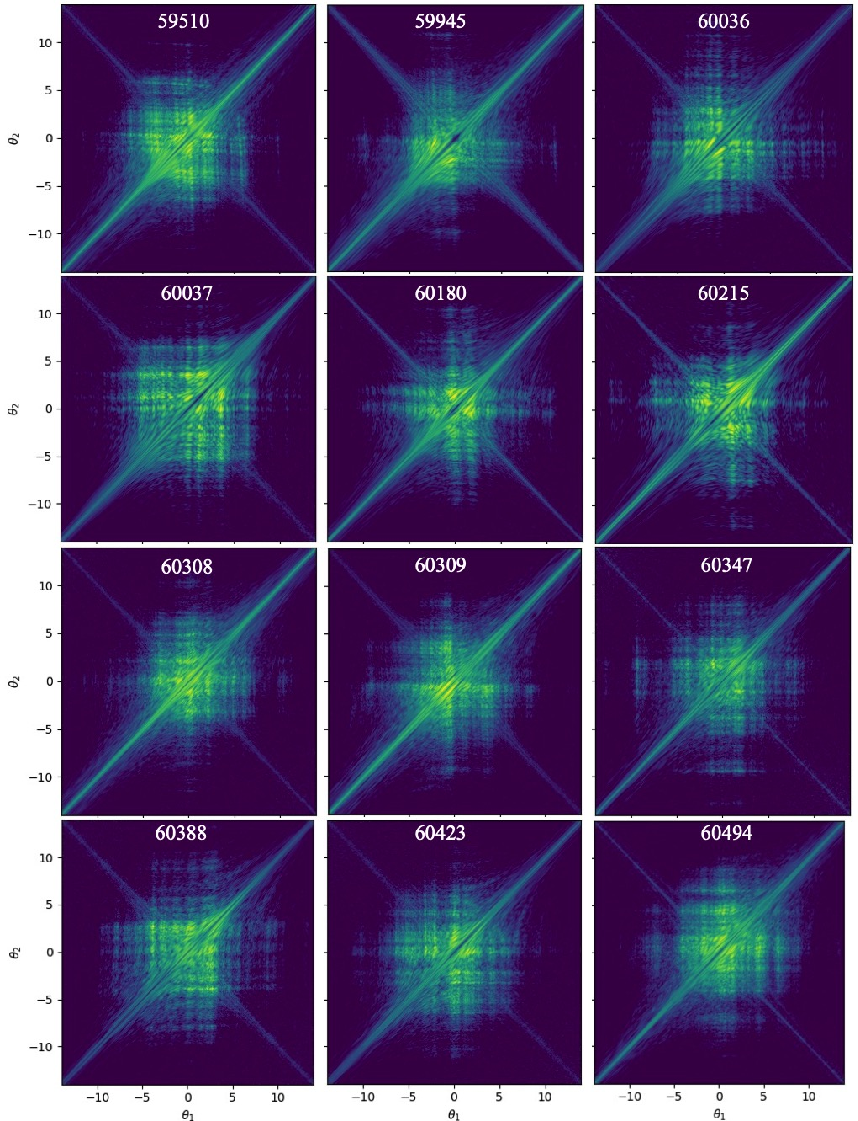}
    \caption{The $\theta$-$\theta$ map for PSR~J1740+1000 from twelve FAST observations centered at 1100~MHz.}
\label{fig:1100_fit}
\end{figure}

\begin{figure}[!ht]
\center
 \includegraphics[width=8.0 cm, angle=0]{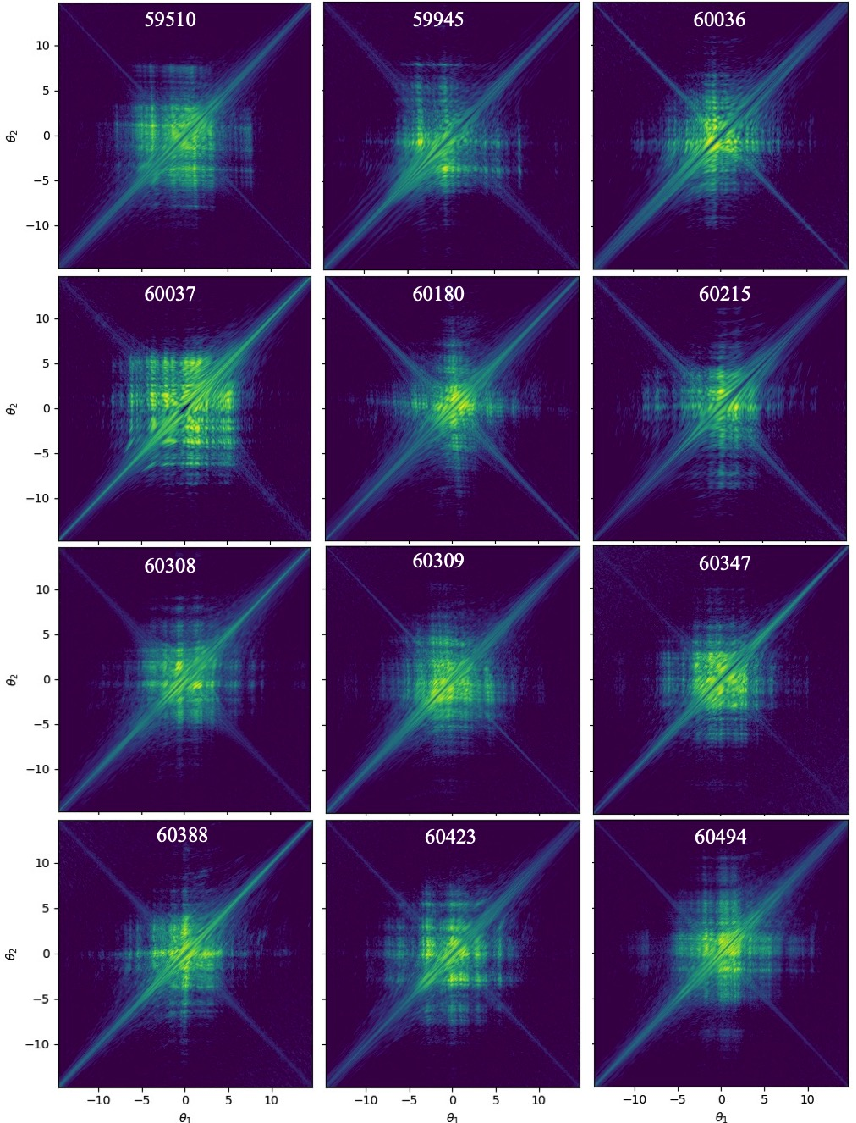}
    \caption{The $\theta$-$\theta$ map for PSR~J1740+1000 from twelve FAST observations centered at 1375~MHz.}
\label{fig:1375_fit}
\end{figure}

\begin{figure}[!ht]
\center
 \includegraphics[width=6.0 cm, angle=0]{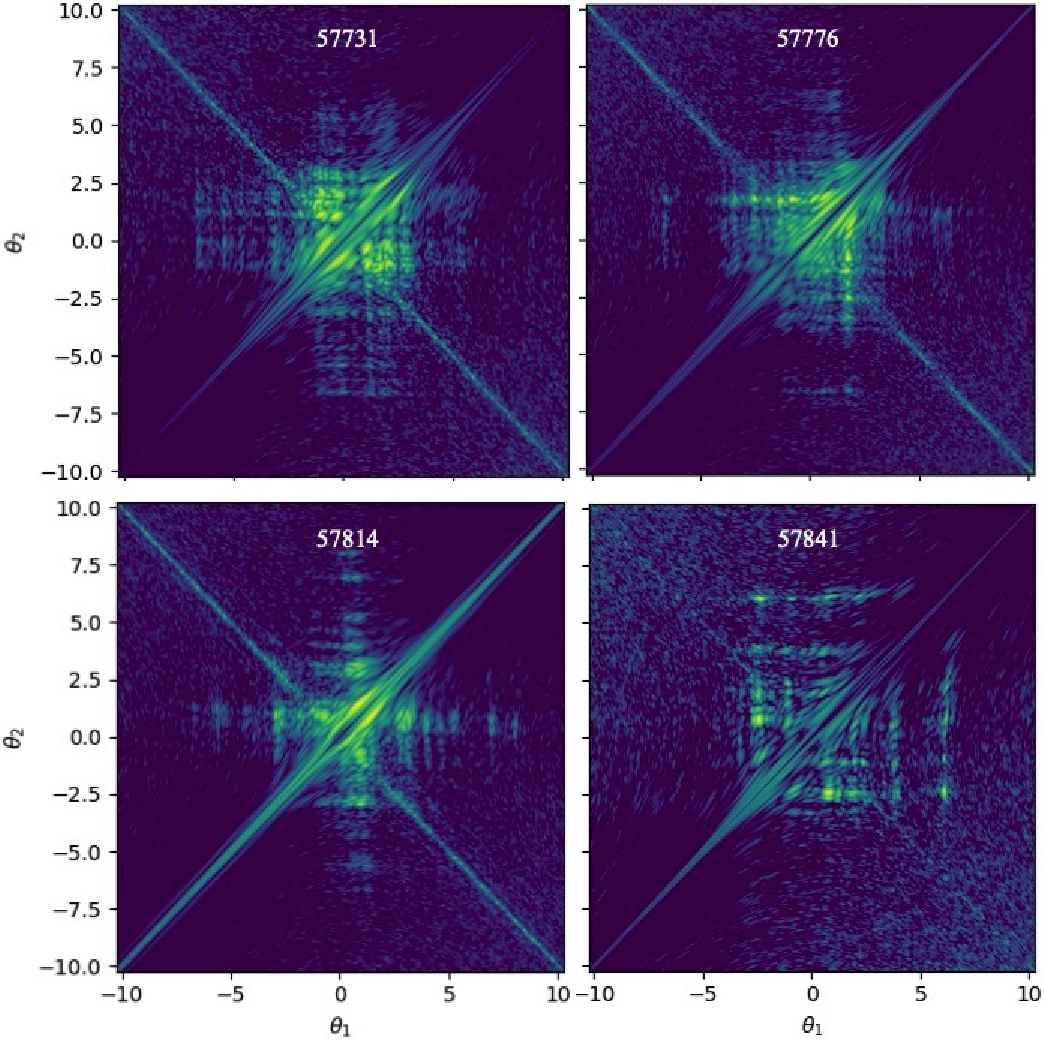}
    \caption{The $\theta$-$\theta$ map for PSR~J1740+1000 from four Parkes observations centered at 1369~MHz.}
\label{fig:1369_fit}
\end{figure}

In the secondary spectrum, a parabolic arc can be described by using
\begin{equation}
f_\nu=\eta f_t^2
\end{equation}
where $\eta$ is the arc curvature. The arclets often have the same curvature as the primary arc. To estimate the curvature of these arclets, we follow the method of \cite{swm+21} and \cite{bbk+22}. Using ththmod.py from scintool \footnote{\url{https://github.com/danielreardon/scintools/tree/master/scintools}}, we first define an appropriate curvature range and then transform the secondary spectra from $f_\nu$-$f_t$ space to $\theta$-$\theta$ space for different curvature values within this range. Each curvature transformation corresponds to a largest eigenvalue in the $\theta$-$\theta$ space. By fitting the peak region of the largest-eigenvalue distribution as a function of curvature with a parabolic function, as shown in Figures~\ref{fig:1100_max}, \ref{fig:1400_max} and \ref{fig:1369_max} in Appendix~\ref{sec:LED}, we determine the best-fit curvature, which are listed in the fourth and fifth columns of Table~\ref{tab:arc_curvature}. From Figures~\ref{fig:1100_max}, \ref{fig:1400_max}, and \ref{fig:1369_max}, compared with the other fits, the 1400~MHz cases for MJDs~60180, 60308, 60309 and the 1369~MHz case for MJD~57814 follow the overall trend of the largest eigenvalue well and yield stable curvature estimates. However, beyond the peak value, the relatively flat behavior of the eigenvalue at larger $\eta$ indicates that the corresponding uncertainties may be underestimated. As shown in Figures~\ref{fig:1100_fit}, \ref{fig:1375_fit} and \ref{fig:1369_fit}, with the best-fit curvature, in the $\theta$-$\theta$ space, where $\theta_1$ and $\theta_2$ are scaled sky coordinates at which the images would interfere at the given $f_\nu$ and $f_t$, all these arclets are concentrated along straight lines that are aligned parallel to the axes suggesting that, for each epoch, the arclets are coming from same scattering screen. 

\begin{table*}[!ht]
\caption{Values and uncertainties of the arc curvature $\eta$ and the distance-weighted effective velocity W for PSR~J1740+1000.}
\centering
\setlength\tabcolsep{8.0pt}
\footnotesize
%\resizebox{\textwidth}{16mm}{
\begin{tabular}{ccccccc}% four columns, alignment for each
	\hline
	MJD& Day of year&Time & $\eta_{1.1}$ & $\eta_{1.375/1.369}$ & W \\
	\hline
	& & min &s$^3$ & s$^3$ &  km~s$^{-1}$~kpc$^{-0.5}$\\
	\hline
    57731& 344& 62 & & 0.0108$\pm$0.0003 &478.01$\pm$6.64\\
    57776& 23& 68 & & 0.0121$\pm$0.0003 &451.61$\pm$5.60\\
    57814& 61& 61 & & 0.0125$\pm$0.0004 &444.32$\pm$7.11\\
    57841& 88& 59 & & 0.0110$\pm$0.0003 &473.65$\pm$6.46\\
	59510&296&118& 0.0196$\pm$0.0004 & 0.0125$\pm$0.0002&  441.99$\pm$4.02\\
	59945&1& 57 & 0.0204$\pm$0.0005 & 0.0128$\pm$0.0003&   435.01$\pm$5.21\\
	60036& 92&48 & 0.0188$\pm$0.0003 & 0.0122$\pm$0.0002&  449.35$\pm$3.63\\
	60037& 93& 48 & 0.0186$\pm$0.0003 & 0.0122$\pm$0.0002& 450.56$\pm$3.66\\
	60180& 236& 58 & 0.0201$\pm$0.0004 & 0.0134$\pm$0.0005&431.67$\pm$6.16\\
    60215& 271& 38 & 0.0190$\pm$0.0003 & 0.0115$\pm$0.0002&454.87$\pm$3.78\\
	60308&365& 118& 0.0189$\pm$0.0006 & 0.0120$\pm$0.0004& 450.61$\pm$7.33\\
	60309& 1& 118& 0.0193$\pm$0.0005 & 0.0129$\pm$0.0003&  440.25$\pm$5.41\\
	60347& 38& 118& 0.0191$\pm$0.0005 & 0.0124$\pm$0.0003& 445.76$\pm$5.61\\
	60388& 79&118& 0.0181$\pm$0.0006 & 0.0120$\pm$0.0003&  455.52$\pm$6.63\\
    60423& 114& 118& 0.0191$\pm$0.0005 & 0.0123$\pm$0.0003&446.66$\pm$5.65\\
	60494& 185&178& 0.0193$\pm$0.0006 & 0.0124$\pm$0.0003& 444.59$\pm$6.15\\ 
	\hline
	\end{tabular}
	\label{tab:arc_curvature}
\end{table*}

\subsection{The location of the scattering screen}\label{sec:location} 
From \cite{mnc+23}, the arc curvature is
\begin{equation}
\eta=4625\frac{D_{\rm eff, kpc}}{\nu^2_{\rm GHz}V_{\rm eff, km/s}^2\rm cos^2\alpha }\label{eq:arc_c}  
\end{equation}
where $\eta$ is in $s^{3}$, $\nu$ is the band-centre frequency, and $\alpha$ is the angle between $\vec{V}_{\rm eff}$ and the direction of the anisotropy axis in the scattered image. The expressions for $D_{\rm eff}$ and $\vec{V}_{\rm eff}$ are
\begin{equation}
 D_{\rm eff}=(1/s-1)D
\end{equation}
where D is pulsar distance, and the scattering screen is located at a distance sD from the pulsar; and
\begin{equation}
\vec{V}_{\rm eff} = (1/s-1) \vec{V}_{\rm pulsar}  + \vec{V}_{{\rm Earth}} - (1/s)\vec{V}_{\rm scr},  
\end{equation}
where $\vec{V}_{\rm pulsar}$ is the pulsar transverse velocity obtained from the measured proper motion, $\vec{V}_{{\rm Earth}}$ is the component of the Earth’s velocity perpendicular to the line of sight to the pulsar and $\vec{V}_{\rm scr}$ is the component of the velocity of the scattering screen perpendicular to the line of sight.

\begin{figure}[!ht]
\center
 \includegraphics[width=6.0 cm, angle=270]{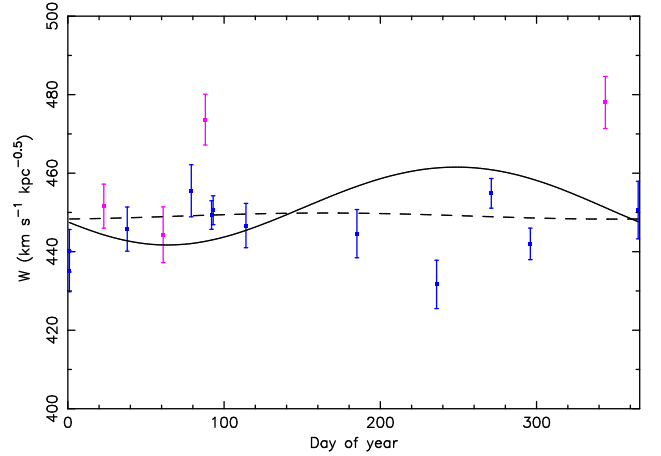}
    \caption{Distance-weighted effective velocity W plotted against day of year. Blue dots correspond to FAST observations, while pink dots represent Parkes measurements. The solid and dashed lines show the best-fit results for the isotropic and anisotropic cases, respectively.}
\label{fig:arc_Earth}
\end{figure}

To make use of all available measurements to estimate the location of the scattering screen, we follow \cite{mnc+23} and analyze the distance-weighted effective velocity W. From Equation~\ref{eq:arc_c}, the frequency-independent form of W is defined as
\begin{equation}
W=\frac{|V_{\rm eff, \parallel}|}{\sqrt{D_{\rm eff}}} = \sqrt{\frac{4625}{\nu^2_{\rm GHz}\eta}}
\label{eq:W}
\end{equation}
where $V_{\rm eff, \parallel}$ is the effective velocity component parallel to the anisotropy axis. For the FAST measurements of $\eta_{1.1}$ and $\eta_{1.375}$, we first compute the corresponding values of W and then take their average. For the Parkes observations, we likewise compute the associated W values. The resulting W for all 16 epochs are listed in the sixth column of Table~\ref{tab:arc_curvature}. These measurements from Parkes and FAST observations span 13 distinct phases of Earth’s orbit, providing good coverage of the full orbital cycle. To investigate whether the arc curvature of PSR~J1740+1000 is modulated by Earth's annual motion, we plot W as a function of day of year in Figure~\ref{fig:arc_Earth}. 

By neglecting the screen velocity and using Equation~\ref{eq:W} along with the measured W values, we determine the screen location under two scenarios: isotropic scattering and anisotropic scattering. Under isotropic scattering, $|V_{\rm eff, \parallel}|$ is given by
\begin{equation}
|V_{\rm eff, \parallel}|=\left|(1/s-1)\vec{V}_{\rm pulsar}  + \vec{V}_{{\rm Earth}}\right|.
\label{eq:V_I}
\end{equation}
Under anisotropic scattering, following Equation~15 of \cite{mnc+23}, it is given by
\begin{multline}
|V_{\rm eff, \parallel}| = \Big| 
  \big(\frac{1-s}{s} V_{\rm pulsar,\alpha} + V_{\rm Earth,\alpha} \big) \sin\psi \\
  + \big(\frac{1-s}{s} V_{\rm pulsar,\delta} + V_{\rm Earth,\delta} \big) \cos\psi
\Big|
\label{eq:V_II}
\end{multline}
where the anisotropy axis is oriented at an angle $\psi$ measured from the positive declination direction. Based on the YMW16 model, the dispersion measure (DM) distance to PSR~J1740+1000 is estimated to be 1.22$\pm$0.24~kpc. Using the newly measured proper motion, the transverse velocity of PSR~J1740+1000 is found to be 329$\pm$80~km/s. As shown in Table~\ref{tab:s_Dsp}, using the pulsar distance and proper motion, we obtain the following: under isotropic scattering, the value of s is derived by fitting the variation of W over the day a year; under anisotropic scattering, both s and the anisotropy angle $\psi$ are determined. In Figure~\ref{fig:arc_Earth}, the solid line represents the best-fit model for isotropic scattering, while the dashed line shows the corresponding fit for the anisotropic case. Based on the values of s and the pulsar distance, we obtain a screen–pulsar distance of 370$\pm$72~pc for isotropic scattering and 1$\pm$12~pc for anisotropic scattering.

From the sixth column of Table~\ref{tab:arc_curvature}, the values of W show little variation, with a standard deviation of 12~km~s$^{-1}$~kpc$^{-0.5}$, which is comparable to the 2$\sigma$ uncertainty of W. As shown by the best-fit results for both isotropic and anisotropic scattering in Figure~\ref{fig:arc_Earth}, the observed variation in W does not originate from modulation by Earth’s motion. Under the isotropic fit, the trend of W varies in the opposite sense to that expected from Earth’s orbital motion. One might suspect that this opposite trend arises because $\vec{V}_{\rm eff}$ is oriented at an angle relative to the anisotropy axis, i.e., due to anisotropic scattering. However, under the anisotropic fit, the inferred scattering screen lies very close to the pulsar, making modulation by Earth even less likely. Therefore, the observed variation in W is most likely attributable to measurement uncertainties and to changes in the magnitude and direction of the screen velocity. From Figure~3 of \cite{kmp+08}, aside from the extended PWN tail, the overall size of the nebula is roughly 2$^\prime$, corresponding to a physical scale of about 0.7~pc. From Table~\ref{tab:s_Dsp}, the reduced chi-square value obtained under anisotropic scattering is about 60\% of that for the isotropic case. Moreover, the screen location inferred from the anisotropic fit is consistent with the extent of the PWN. Therefore, for PSR~J1740+1000, it is highly likely that the scattering screen is associated with the PWN.

In practice, we also attempted to include the screen velocity in both the isotropic and anisotropic models. However, introducing this additional parameter not only failed to reduce the reduced chi-square value, but also led to fitted parameters that were not physically reasonable. From recent pulsar scintillation studies \citep{rch+19, mnc+23}, the screen velocity is typically in the range of 10–20~km/s. Since the scattering screen in these two cases is located close to the pulsar and the screen velocity is much smaller than the pulsar’s transverse velocity, the screen velocity has a negligible effect on our results.

\begin{table}
\caption{Values and uncertainties of s, $\psi$, $\chi^2_\nu$ and $D_{\rm sp}$ under isotropic and anisotropic scattering.}
\centering
\setlength\tabcolsep{4.0pt}
\footnotesize
\begin{tabular}{ccc}
	\hline
	   Parameters& Isotropic& Anisotropic \\
	\hline
	s&0.303$\pm$0.003& 0.0008$\pm$0.0097\\
$\Psi$~(deg.)& & 96.1$\pm$14.6\\
$\chi^2_\nu$ & 7.13 & 4.25\\
$D_{\rm sp}$~(pc)& 370$\pm$72& 1$\pm$12\\
	\hline
	\end{tabular}
	\label{tab:s_Dsp}
\end{table}

\subsection{The double-layered adjacent arcs and arclets}\label{sec:double_arcs}
In Figure~\ref{fig:ss_1100} and Figure~\ref{fig:ss_1375}, all secondary spectra display reverse arclets with several showing double or even triple-layered adjacent arcs, notably on MJDs~60036, 60037, 60180, and others. Among these, the secondary spectrum on MJD~60180 at 1100~MHz exhibits the most pronounced double-layered adjacent arcs and comparatively weaker reverse arclets, which makes it particularly well suited for analyzing each arc layer individually. To better study the frequency dependence of the corresponding angular positions of the sub-images responsible for the double-layered adjacent arcs, we select the secondary spectra on MJD~60180 at both 1100 (1050-1150~MHz) and 1400~MHz (1350-1450~MHz) for the detailed analysis presented below.

\begin{figure}[!h]
\center
 \includegraphics[width=8.0 cm, angle=270]{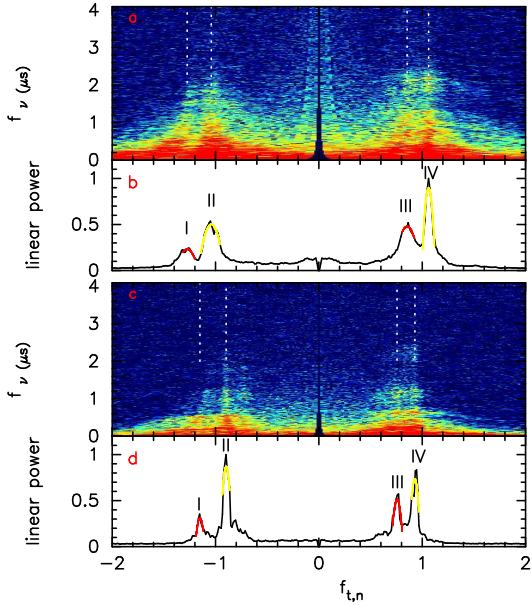}
    \caption{The normalized secondary spectra and their corresponding Doppler profiles at 1100~MHz (panels a and b) and 1400~MHz (panels c and d). Panels a and c: the white dashed lines show the best-fit $\beta$ for each peak. Panels b and d: the red and yellow lines represent parabolic fits to the arc peaks.} 
\label{fig:fit_60180}
\end{figure}

Initially, with the true curvatures unknown, we assumed that the double-layered adjacent arcs are centered on the coordinate origin but possess different curvatures. Following \cite{rcb+20}, to estimate the arc curvature of these double-layered adjacent arcs separately, we normalized the secondary spectra at 1100 and 1400~MHz with respect to an arc curvature of $\eta_0=0.02~\rm s^3$ and $\eta_0=0.009~\rm s^3$, respectively. This normalization ensures that an arc with curvature $\eta_0/\beta^2$, with no shift relative to the coordinate center, appears as two vertical lines in the normalized secondary spectrum at $f_{t,n}=\beta_+$ and $f_{t,n}=\beta_-$, as shown in panels (a) and (c) of Figure~\ref{fig:fit_60180}. To optimally estimate the arc curvature, we averaged the power along the $f_\nu$ axis to form the power distribution as a function of $f_{t,n}$, which we refer to as the ``Doppler profile”. To minimize the effect of strong signals at small $f_\nu$, we averaged the power only for $f_\nu>$0.5~$\mu$s. We then normalized the $f_{t,n}$ axis so that the value of the highest peak is 1.0. The corresponding Doppler profiles at 1100 and 1400~MHz exhibit four distinct peaks (I, II, III and IV), as shown in panels (b) and (d) of Figure~\ref{fig:fit_60180}. After that, depending the peak width, we selected three to five data points on either side of the peak to define the peak region. Finally, for each peak, we fitted the peak region with a parabolic function, and the resulting values of $\beta$ are listed in the second and third columns of Table~\ref{tab:beta_gamma}. In Panels (b) and (d) of Figure~\ref{fig:fit_60180}, the red and yellow lines represent the best-fit results. From Table~\ref{tab:beta_gamma} and Figure~\ref{fig:fit_60180}, we note that the values corresponding to the two innermost peaks ($\beta_{II}$ and $\beta_{III}$) and the two outermost peaks ($\beta_I$ and $\beta_{IV}$) are not equal at both 1100 and 1400~MHz, indicating that they do not correspond to arcs centered on the coordinate origin with different curvatures. Instead, we find that the mean value of $\beta_I$ and $\beta_{III}$ is consistent with that of $\beta_{II}$ and $\beta_{IV}$ at both 1100 and 1400~MHz. Furthermore, as shown in Figure~\ref{fig:1100_fit}, the two arcs on MJD~60180 at 1100~MHz appear parallel when transformed from the $f_\nu$–$f_t$ space to the $\theta$–$\theta$ space using the same arc curvature of $\eta_{1.1}=0.0201\pm0.0004$~s$^3$. Based on these two pieces of evidence, we conclude that the double-layered adjacent arcs with the same curvature originate from the same scattering screen, with one being the primary arc and the other a strong arclet. 

\begin{figure}[!h]
\center
 \includegraphics[width=8.5 cm, angle=270]{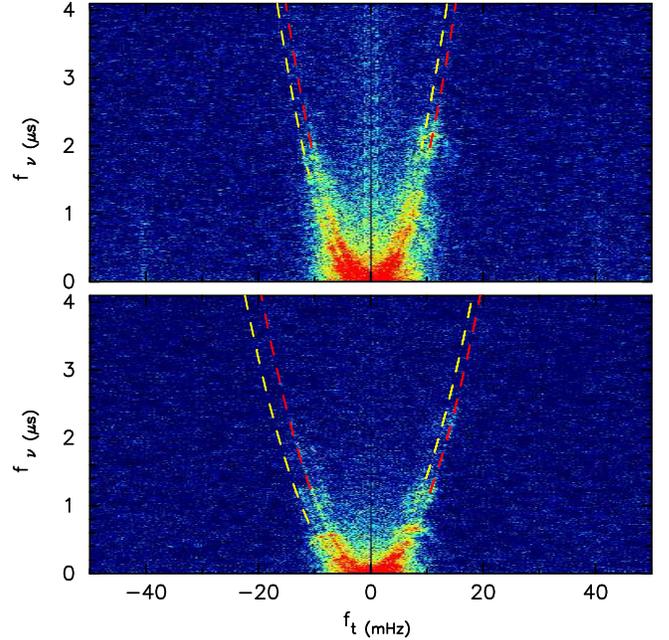}
    \caption{The secondary spectra on MJD~60180 at 1100 and 1400~MHz. The yellow and red dashed lines represent the best-fit arc curvatures.}
\label{fig:DP}
\end{figure}

According to \cite{hsa+05} and \cite{crs+06}, both the primary arc and the strong arclet can be described by the equation:
\begin{equation}
f_\nu=\eta f_t^2 +\gamma f_t
\label{eq:shift}
\end{equation}
where $\eta$ is the arc curvature as shown in Equation~\ref{eq:arc_c}, $\gamma$ is 0 for the primary arc, and for the strong arclet, assuming the screen velocity is negligible, $\gamma$ (in units of s$^2$) is given by
\begin{equation}
    \gamma=0.1496\displaystyle\frac{D_{\rm eff}\theta_{\rm g, mas}}{\nu_{\rm GHz} |V_{\rm eff, \parallel}|}
 \label{eq:gamma}
\end{equation}
where the sub-image responsible for the strong arclets is located at $\theta_{\rm g, mas}$. Solving for $\theta_{\rm g, mas}$ gives:
\begin{equation}
  \theta_{\rm g, mas}=\frac{\gamma \nu_{\rm GHz} |V_{\rm eff, \parallel}|}{0.1496 D_{\rm eff}}, 
 \label{eq:theta_g}
\end{equation}
where $|V_{\rm eff, \parallel}|$ is given by Equation~\ref{eq:V_I} for isotropic scattering, and by Equation~\ref{eq:V_II} for anisotropic scattering. To estimate $\eta_{I-III}$, $\gamma_{I-III}$, $\eta_{II-VI}$ and $\gamma_{II-VI}$ at 1100 and 1400~MHz based on these values of $\beta$, we follow the procedures outlined below. First, we separately calculate the $\eta_0/\beta^2$, i.e. $\eta_I$, $\eta_{II}$, $\eta_{III}$, $\eta_{VI}$, based on values of $\beta$ for I, II, III and IV. Second, based on $\eta_I$ and $\eta_{III}$, and $\eta_{II}$ and $\eta_{VI}$, we produce the left and right sides of two arcs in the secondary spectra with $f_\nu<2.0~\rm\mu s$. Third, we fit the arc generated from $\eta_I$ and $\eta_{III}$ to obtain $\eta_{I-III}$ and $\gamma_{I-III}$, and fit the arc generated from $\eta_{II}$ and $\eta_{VI}$ to obtain $\eta_{II-VI}$ and $\gamma_{II-VI}$. In Table~\ref{tab:beta_gamma}, we show the value of these parameters with 1$\sigma$ uncertainty. In Figure~\ref{fig:DP}, the yellow and red dashed lines represent the best-fit results. From the left sides of the secondary spectra at both 1100 and 1400~MHz, where the double-layered adjacent arcs have a larger separation, we can see that the fitting results are quite good. In fact, during our analysis we generated the Doppler profiles for all observing epochs. However, we found that, apart from MJD~60180, it is not possible to extract the corresponding $\eta$ and $\gamma$ values for each arc layer from these Doppler profiles, owing to the contaminating influence of the reverse arclets. From \cite{hsa+05} and \cite{rsz+21}, for discrete reverse arclets, if their angular positions are independent of frequency, they are interpreted as arising from a discrete AU-scale structure within the scattering region. As noted above, the double-layered adjacent arcs originate from the same scattering screen, with one corresponding to the primary arc and the other to a discrete, strong arclet. From Equation~\ref{eq:gamma}, if the refraction of the strong arclet is produced by such an AU-scale structure and the angular position of the corresponding sub-image is independent of observing frequency, then the ratio of $\gamma_{I-III}$ at 1.1~GHz to $\gamma_{I-III}$ at 1.4~GHz should be 1.4/1.1$=$1.273, regardless of whether the scattering is isotropic or anisotropic. From Table~\ref{tab:beta_gamma}, we find that the ratio of $\gamma_{I-III}$ at 1.1~GHz to $\gamma_{I-III}$ at 1.4~GHz is $1.277\pm0.009$, which is consistent with 1.273 within 0.5$\sigma$, suggesting that the refraction of the strong arclet comes from an AU-scale structure located within the scattering region. Following \cite{ycm+22}, using Equations~\ref{eq:gamma} and \ref{eq:theta_g}, along with the measured proper motion, $\eta$, $\gamma$, the DM-based distance and $\psi$ measured in Section~\ref{sec:location}, we obtain the value and uncertainty of $\theta_g$ at 1100 and 1400~MHz for both isotropic scattering ($\theta_{g1}$) and anisotropic scattering ($\theta_{g2}$). From Table~\ref{tab:beta_gamma}, in the isotropic scattering case, the $\theta_{g1}$ values derived from the two frequency bands are identical to each other. The same is true for the anisotropic scattering case.

In Section~\ref{sec:ss_arc}, the measured arc curvature on MJD~60180 at 1100~MHz using the $\theta$-$\theta$ method is 0.0201$\pm$0.0004~s$^3$, which differs by 6$\sigma$ from the value of 0.0181$\pm$0.0002~s$^3$ given in Table~\ref{tab:beta_gamma}. The difference between these two values is primarily due to the different focuses of the two methods. The former yields the average curvature of all detected arclets, while the latter provides the curvature for the primary arc and the strong arclet. In the secondary spectra, for the continuous reverse arclets detected in each epoch, due to their relatively diffuse nature, we cannot study their frequency dependence. Therefore, it is difficult to determine their origin mechanism. They could either arise from stochastic, speckle-like substructures in the scattered image due to a stochastic screen with a Kolmogorov spectrum, or from discrete features in the scattered image due to refraction by a discrete, dense AU-scale structure. 
\begin{table}
\caption{Value and uncertainty of $\beta$, $\eta$, $\gamma$ and $\theta_g$ for the double-layered adjacent arcs on MJD~60180 at 1100 and 1400~MHz.}
\centering
\setlength\tabcolsep{4.0pt}
\footnotesize
\begin{tabular}{ccc}
	\hline
	   & 1100~MHz& 1400~MHz\\
	\hline
	$\beta_{I}$&$-$1.274$\pm$0.016&$-$1.151$\pm$0.002\\
	$\beta_{II}$&$-$1.041$\pm$0.008&$-$0.897$\pm$0.002\\
	$\beta_{III}$&0.854$\pm$0.009&0.756$\pm$0.003\\
	$\beta_{IV}$&1.062$\pm$0.005&0.929$\pm$0.003\\
	$\eta_{I-III}$ (s$^3$)& 0.0179$\pm$0.0003& 0.0101$\pm$0.0001\\
	$\gamma_{I-III}$ (s$^2$)&0.0553$\pm$0.0003& 0.0433$\pm$0.0002\\
	$\eta_{II-VI}$ (s$^3$)&0.0181$\pm$0.0002& 0.0108$\pm$0.0001\\
    $\gamma_{III-VI}$ (s$^2$)& 0&0\\
    $\theta_{g1}$ ($\mu$as) & 112$\pm$16 & 112$\pm$16\\
     $\theta_{g2}$ ($\mu$as) & 23$\pm$17 & 23$\pm$17\\
	\hline
	\end{tabular}
	\label{tab:beta_gamma}
\end{table}

\begin{figure}[!h]
\center
 \includegraphics[width=8 cm, angle=0]{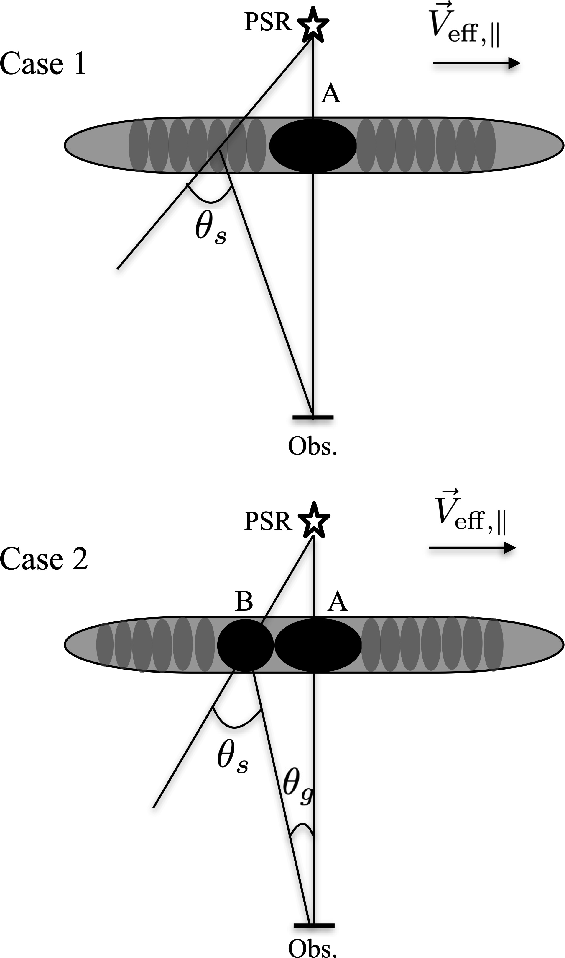}
    \caption{Cartoon of pulsar scattered images. Top: A single dominant scattering image produces one main arc with associated arclets. Bottom: Two dominant scattered images result in two main arcs, each with its own arclets. $\vec{V}_{\rm eff, \parallel}$ denotes the effective velocity component parallel to the anisotropy axis of the scattered image.}
\label{fig:model}
\end{figure}

In Figure~\ref{fig:model}, we present a cartoon illustrating the pulsar’s scattered images under two scenarios, aimed at interpreting the observed arclets and the double-layered adjacent arcs in PSR~J1740+1000. In both cases, the scattered images appear as approximately linear structures, with $\vec{V}_{\rm eff, \parallel}$ aligned along the direction of the images. In Case 1, the pulsar image consists of a single dominant unscattered component (A) along with multiple weaker scattered components. In contrast, Case 2 includes one additional dominant scattered component (B) situated near component A. Following the simulation work on the origin of arcs and arclets by \citet{crs+06}, in Case 1, interference between component A and the remaining scattered components produces a single dominant main arc. Additionally, interference between one of the weaker scattered components and the rest of the image gives rise to arclets. Thus, Case 1 provides a plausible explanation for the presence of a single dominant arc with multiple arclets observed in many pulsars, for example, PSR~B1133+16 \citep{tr07}. In Case 2, interference between component B located at $\theta_g$ and the remaining components produces an additional dominant arc, or equivalently, a strong arclet. This scenario naturally explains the simultaneous appearance of arclets and multiple arcs in PSR~J1740+1000 across multiple epochs. 

In Section~\ref{sec:DS}, we note that if the scattered pulsar image is dominated by two sub-images, as observed on MJD~60180, the modulation in the 1D ACF is more pronounced in the frequency domain than in the time domain. From Equations 2 and 3 of \cite{crs+06}, the modulation periods in time and frequency are as follows,
\begin{equation}
T_t=\frac{\lambda s}{\theta_g\left|V_{\rm eff, \parallel}\right|}
\end{equation}
and 
\begin{equation}
T_\nu=\frac{2cs}{\theta^2_g(1-s)D}.
\end{equation}
For dynamic spectra, the scintillation timescale ($\Delta t_d$) and the scintillation bandwidth ($\Delta \nu_d$), are as follows:
\begin{equation}
\Delta t_d=\frac{\lambda}{2\pi\theta_d \left|V_{\rm eff, \parallel}\right|}
\end{equation}
and 
\begin{equation}
\Delta \nu_d=\frac{1}{2\pi\tau_s}=\frac{c}{\pi\theta^2_d(1-s)D}
\end{equation}
where $\theta_d$ is the width of the angular spectrum or the angular scale of the Component A and $\tau_s$ is the scattering timescale. From these equations, we find that 
\begin{equation}
\frac{\Delta t_d}{T_t}=\frac{1}{2\pi s}\frac{\theta_g}{\theta_d}
\end{equation} 
and 
\begin{equation}
\frac{\Delta \nu_d}{T_\nu}=\frac{1}{2\pi s}\frac{\theta^2_g}{\theta^2_d}. 
\end{equation} 
As $\theta_g>\theta_d$, the ratio $\frac{\Delta t_d}{T_t}$ is smaller than $\frac{\Delta \nu_d}{T_\nu}$, thus the periodic modulation in the frequency domain is more pronounced.

\section{Summary and Conclusions}\label{sec:SC}
By combining observations from the Nanshan, FAST, and Parkes telescopes, we conducted a timing and scintillation study of PSR~J1740+1000. Through timing analysis, we detected a large glitch and measured the pulsar’s proper motion for the first time. Using a scintillation analysis, along with the measured proper motion and DM-based distance, we investigated the properties and constrained the screen location for both isotropic and anisotropic scattering cases. Additionally, at several epochs we simultaneously detect reverse arclets together with double- or even triple-layered adjacent arcs.

The glitch, with a magnitude of 2899.2(8)$\times$10$^{-9}$, detected in Nanshan data on 56167(19.5) closely  matches the large glitch observed in the Jodrell Bank data. The derived proper motion of $\mu_\alpha=$8.5$\pm$3.2~mas/yr and $\mu_\delta=$56.3$\pm$8.0~mas/yr indicates that PSR~J1740+1000 is a typical-velocity pulsar, with a total proper motion of $\mu_{\rm tot}=56.9\pm8.0$~mas/yr and a transverse velocity of V$_t=329\pm80$~km/s. The motion is directed away from the Galactic plane at an angle of 16$^\circ.7\pm4^\circ.8$. Based on Chandra observations, \cite{hbg13} placed an upper limit of 60~mas/yr on the proper motion of PSR~J1740+1000. Our measured proper motion is consistent with this limit. From the detection of diffuse PWN tail in X-ray, \cite{kmp+08} inferred that the pulsar is moving toward the Galactic plane at a very small angle of 7$^\circ$, though this rough estimate was made without quoting uncertainties. The two results differ by approximately 23$^\circ$. As mentioned by \cite{cbc+11}, fitting the position using higher-order spin parameters in timing tends to underestimate the uncertainty in the obtained position, which could partly explain the discrepancy in the results. In the future, high-precision timing or high-resolution X-ray observations will be needed to cross-check these findings. Based on the newly measured proper motion and its DM-distance, we concluded that PSR~J1740+1000 is a young pulsar born in the Galactic halo. Similar to the case of PSR~J0837$-$2454, another young pulsar located more than 1~kpc above the Galactic plane, as discussed by \citet{Pol_2021}, PSR~J1740+1000 may also have originated from a runaway O or B-type star. By combining the peculiar velocities and average lifetimes of runaway stars, \citet{Pol_2021} found that such stars could travel distances of up to ~2 kpc before collapsing into neutron stars. This provides a natural explanation for the large vertical (z) heights of these two young halo pulsars.

From FAST and Parkes observations, we detected scintillation arcs, reverse arclets and double-layered adjacent arcs for PSR~J1740+1000. The dynamic spectra exhibit a periodic fringe or criss-cross pattern, where the pulsar's scattered image is dominated by several sub-images. For several epochs, the pulsar's scattered image is dominated by two sub-images, and the corresponding time-domain 1D ACF and frequency-domain 1D ACF show clear periodic modulation. The modulation in the frequency domain is more pronounced than in the time domain. From the secondary spectra, for all epochs, we measured the curvature at 1100, 1375, and 1369~MHz from these reverse arclets and found no evidence for Earth's annual modulation, indicating that the scattering screen is located closer to the pulsar. With the new proper motion measurements and the DM-distance, we determined that the scattering screen lies 370$\pm$72~pc from PSR~J1740+1000 under isotropic scattering, and 1$\pm$12~pc from the pulsar under anisotropic scattering. These distances indicate that the scattering screen isn't NPS or Gould Belt, or any other screens located near Earth. In contrast with the isotropic case, the anisotropic model not only yields a smaller reduced chi-square, but also gives a screen location consistent with that of the PWN, making it the more likely scenario. From several epochs, we simultaneously detected reverse arclets together with double- or even triple-layered adjacent arcs. Based on the detection from MJD~60180 at 1100 and 1400~MHz, we find that one is the primary arc and the other is a strong arclet. For the strong arclet, the angular position of the corresponding sub-image is independent of the observing frequency, suggesting that this sub-image originates from refraction by an AU-scale structure located within the scattering region. As noted above, it is highly likely that the PWN dominates the scattering of PSR~J1740+1000. From XMM-Newton X-ray observations, the PWN of PSR~J1740+1000 contains numerous diffuse small-scale structures \citep{kmp+08, bbb+21}. Therefore, these AU-scale structures may be associated with the PWN of PSR~J1740+1000. Our analysis suggests that the scattered image of PSR~J1740+1000 consists of two dominant components: one located at the pulsar's geometric position and the other located 112$\pm$16 and $23\pm17$~$\mu$as away from the central component under the isotropic and anisotropic scattering cases, respectively, along with multiple weaker components. For these reverse arclets, due to their continuity and diffuse nature, we have not obtained the evolution of the angular position of their corresponding scattered sub-images with frequency, which prevents us from revealing their origin.

In the future, multiple long-term observations with FAST will further enhance the precision of proper motion measurements. Meanwhile, VLBI measurements will not only provide accurate proper motion and parallax measurements but also validate our results. Additionally, continuous multi-day observations will aid in identifying the underlying causes of the pulsar's flux variations. 

%%%%%%%%%%%%%%%%%%%%%%%%%%%%%%%%%%%%%%%%%%%%%%%%%%
\section{Acknowledgements} 
%%%%%%%%%%%%%%%%%%%%%%%%%%%%%%%%%%%%%%%%%%%%%%%%%%
We gratefully acknowledge the anonymous reviewers for their thorough review and valuable suggestions, which significantly contributed to enhancing this work. This work is supported by the Major Science and Technology Program of Xinjiang Uygur Autonomous Region (2022A03013-2, 2022A03013-3), the National Science Foundation of Xinjiang Uygur Autonomous Region (2022D01D85), the CAS Project for Young Scientists in Basic Research (YSBR063), the Tianshan talents program (2023TSYCTD0013), the Chinese Academy of Sciences (CAS) “Light of West China” Program (No. xbzgzdsys202410  and No. 2022XBQNXZ015) and the National Key RD Program of China No.  2022YFC2205202. JMY is supported by the Tianchi Talent project. FFK is supported by the open research project funded by the Key Laboratory of Xinjiang Uyghur Autonomous Region (2022D04015), the Natural Science Foundation of China (12203093), the National Key Research and the Development Program (2022YFA1603104). WMY is supported by the National Natural Science Foundation of China (NSFC) project (No. 12273100) and the West Light Foundation of Chinese Academy of Sciences (No. WLFC 2021-XBQNXZ027).

%\clearpage

\begin{appendix}
\section{The glitch of PSR~J1740+1000}\label{sec:glitch}
\begin{figure}[!h]
\center
 \includegraphics[width=8.0 cm, angle=270]{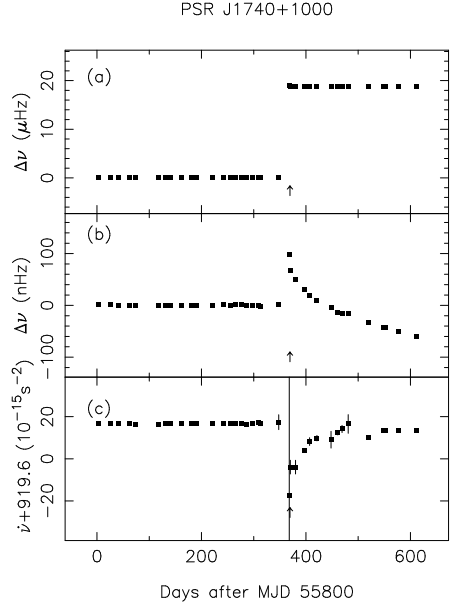}
    \caption{Variations of $\nu$ and $\dot{\nu}$ of PSR~J1740+1000 by fitting $\nu$ and $\dot{\nu}$ for small sections of data. Each section contains about 15 ToAs and repeats for 10 ToAs. (a) Variations of frequency $\Delta \nu$ relative to the pre-glitch solutions; (b) an expanded variation of frequency $\Delta \nu$ relative to the post-glitch solutions; (c) variations of the $\dot{\nu}$ frequency derivative. The arrows of each pannels and the black line in the bottom pannel are the glitch epoch (MJD~56167).}
\label{fig:glitch}
\end{figure}

A glitch was detected in the Nanshan data around MJD~56167, as shown in Figure~\ref{fig:glitch}. In the top panel, a jump in $\nu$ resulting from the glitch is marked by the black arrow. The middle and bottom panels illustrate an exponential recovery process following the glitch. By fitting the pulsar timing/glitch model to the corrected ToAs, we determined the glitch parameters, which are listed in Table~\ref{tab:glitch}, where $\Delta{\nu_{g}}=\Delta{\nu}_{p}+\Delta{\nu}_{d}$. The magnitude of the glitch was determined to be 2899.2(8)$\times10^{-9}$. Additionally, the exponential recovery process exhibits a fractional recovery amplitude of $Q=\frac{\Delta{\nu}_{d}}{\Delta{\nu_{g}}}=0.0032(9)$ and a recovery timescale of $\tau_d=93(17)$ days. As mentioned in Section~\ref{sec:Intro}, two glitches have been detected since the pulsar's discovery. The occurrence time and magnitude of this glitch detected in Nanshan data closely match those of the big glitch detected in the Jodrell Bank data \citep{bsa+22}. After the glitch, the rotational frequency typically undergoes an exponential decay process, indicating that the star is not a single rigid body, but rather consists of a solid crust coupled with a superfluid neutron interior \citep{bpp+69, aap+81}. According to classical glitch theory, the fraction of the super-fluid moment of inertia $I_n/I$ is approximately $5.3\%$ \citep{lbe+99}. However, when taking into account the entrainment, the required moment of inertia increases up to $20\%$ of the total. This suggests that the superfluid within the crust alone is insufficient to explain the observed big glitch, and that the superfluid in the inner core must also contribute to the event as well \citep{cha13}.  
\begin{table}[htb]
\caption{The glitch parameters for PSR~J1740+1000.}
\renewcommand{\arraystretch}{1.25}
\centering
\begin{tabular}{cc}
\hline 
Parameter & Value\\
\hline
Data span (MJD) & 55800-56400 \\
Glitch epoch (MJD) & 56167(19.5) \\
$\Delta{\nu_{g}} (10^{-9}/s^{1})$ & 18814(1)  \\
$\Delta{\nu_{g}}/\nu (10^{-9})$ & 2899.2(8) \\
$\Delta{\dot{\nu_{g}}} (10^{-15}/s^2)$ & $-$2.4(6) \\
$\Delta{\nu_{d}} (10^{-9}/s^{1})$& 62(9) \\
$\tau_{d}$ (day) & 93(17) \\
Q& 0.0032(9)\\
\hline
\end{tabular}
\label{tab:glitch}
\end{table}

\section{Tests of position fitting with higher-order spin-frequency parameter}\label{sec:fit_simulation}
\begin{figure}[!ht]
\center
 \includegraphics[width=10 cm, angle=0]{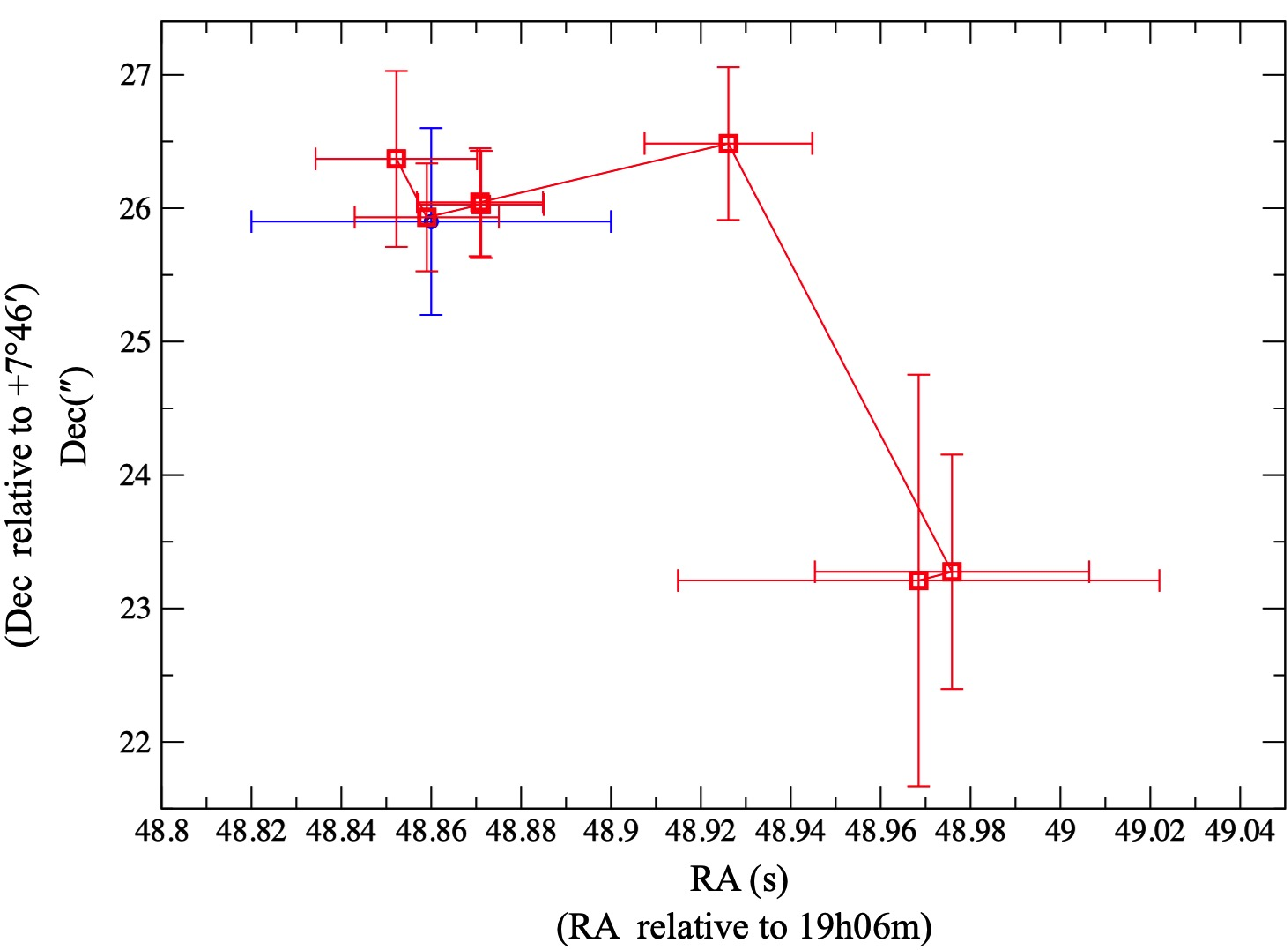}
    \caption{The positions obtained from timing fits of PSR~J1906+0746 using higher-order spin parameters are shown. The blue point marks the true position, while F1, F2, F3, F4, F5, and F6 represent the positions derived when fitting up to the corresponding orders of the spin parameters.}
\label{fig:simulate}
\end{figure}

This appendix presents the validation of our position fitting procedure using PSR~J1906+0746 and its application to PSR~J1740+1000. To assess the impact of timing noise and the inclusion of higher-order spin-frequency derivatives on pulsar position fitting, we performed a dedicated test using PSR~J1906+0746, which has a spin period and period derivative similar to those of PSR~J1740+1000. We obtained the published timing ephemeris for PSR~J1906+0746 from \cite{vks+15}, which includes spin-frequency derivatives up to F4. Using this parameter file (for simplicity, we leave the binary parameters), we generated a set of synthetic times of arrival (TOAs) with the {\tt fake} plugin in TEMPO2. We then modified the right ascension (RA) and declination (Dec) in the parameter file and performed timing fits including different orders of higher-order spin-frequency derivatives to recover the RA and Dec. As illustrated in Figure~\ref{fig:simulate}, once the fit includes derivatives up to F4, the recovered position converges to the true value, and the inclusion of F5 produces negligible changes in both the fitted position and the corresponding timing residuals.

Following this procedure, we applied the same approach to PSR~J1740+1000. For the first data segment, we fitted spin-frequency derivatives up to F4 (upper part of Table~\ref{tab:p_result}), and for the second segment, we fitted up to F3 (lower part of Table~\ref{tab:p_result}). In both cases, the resulting timing residuals are flat, and the uncertainties of the higher-order derivatives are reasonable, confirming that the chosen orders are sufficient to mitigate timing noise without introducing significant bias in the derived positions.
\begin{table}[!ht]
\caption{Fitting results for the first Nanshan dataset (upper part) and the combined Nanshan + FAST dataset (lower part)}
\renewcommand{\arraystretch}{1.25}
\centering
\begin{tabular}{ccc}
\hline 
Parameter& Post-fit Value& Uncertainty \\
\hline
RAJ (hms) &17:40:25.9559&0.0067\\
DECJ (dms)&+10:00:06.74&0.12\\ 
F0 (s$^{-1}$)&6.4895422410806508705& 0.00000000010476836295\\
F1 (s$^{-2}$)&$-$9.0304064606702561806e$^{-13}$& 4.0798621101748564463e$^{-18}$\\
F2 (s$^{-3}$)& 7.0478727135437164105e$^{-24}$&  5.3672238179380352206e$^{-25}$\\
F3 (s$^{-4}$)& 9.675377592062440859e$^{-32}$& 1.3010187552493679451e$^{-32}$\\
F4 (s$^{-5}$)& $-$1.8781268782437589071$^{-38}$& 2.0477960473512576627e$^{-39}$\\
PEPOCH (MJD)& 55384& $/$\\
START (MJD) & 54620.646756&$/$\\
FINISH (MJD)& 56167.000014&$/$\\
\hline
RAJ (hms) &17:40:25.9627&0.0017\\
DECJ (dms)&+10:00:07.47&0.07\\ 
F0 (s$^{-1}$)& 6.4892229783804549355& 0.00000000008982308603\\
F1 (s$^{-2}$)&$-$9.040251100003341224e$^{-13}$& 3.2802620954774010735e$^{-18}$\\
F2 (s$^{-3}$)& 8.2569905566783065646e$^{-25}$&  1.2733473962184035602e$^{-25}$\\
F3 (s$^{-4}$)& $-$3.5210060197480240779e$^{-32}$& 9.5894682234846429662e$^{-33}$\\
PEPOCH (MJD)& 59711& $/$\\
START (MJD) & 58926.871288&$/$\\
FINISH (MJD)& 60495.64819&$/$\\
\hline
\end{tabular}
\label{tab:p_result}
\end{table}

\section{Arc Curvature Measurement from Largest Eigenvalue Distribution}\label{sec:LED}
\begin{figure}[!ht]
\center
 \includegraphics[width=18 cm, angle=0]{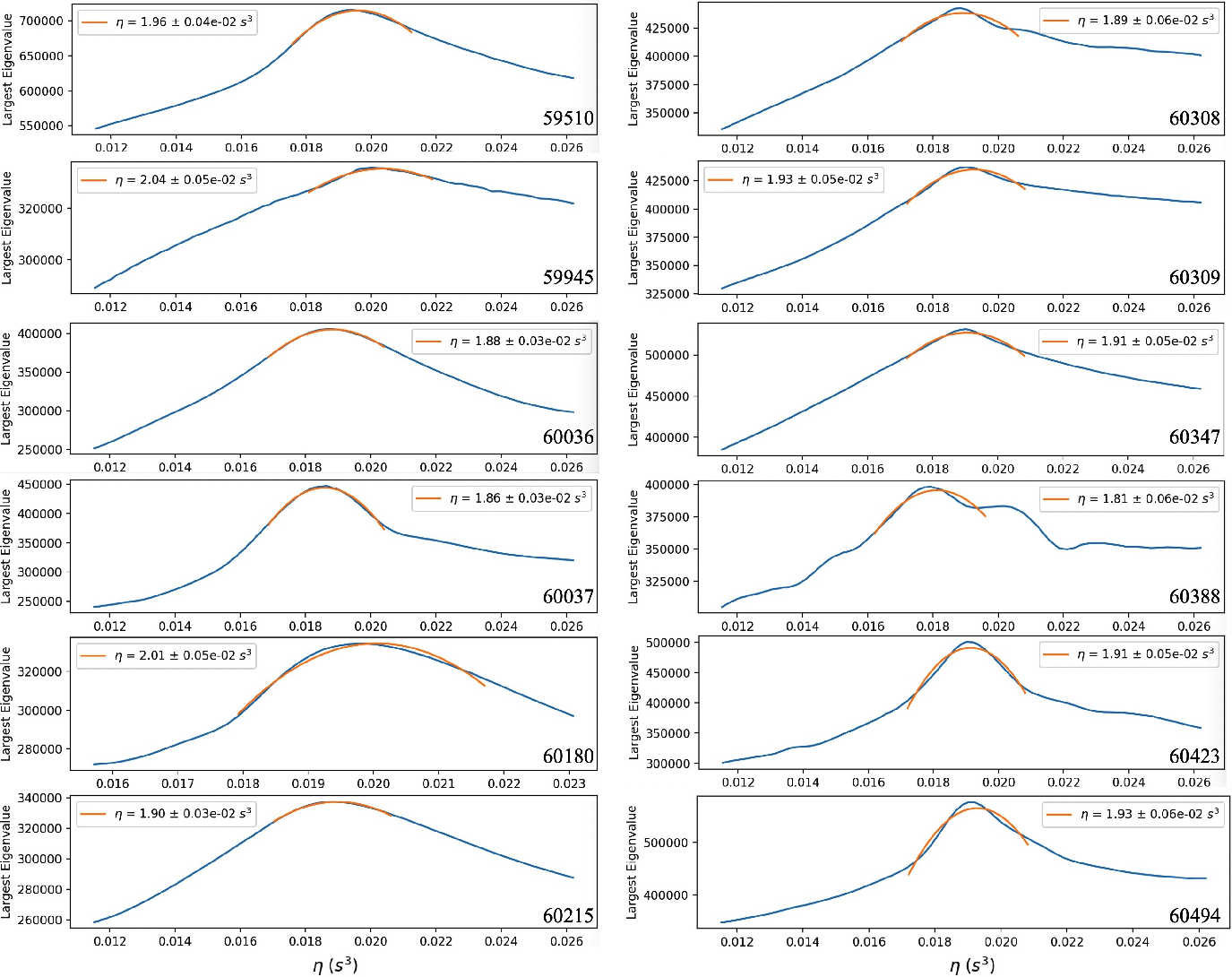}
    \caption{Values of the largest eigenvalue from 12 FAST observations centered at 1100~MHz plotted against arc curvature. The blue lines represent the distribution of eigenvalues evaluated at different arc curvatures, while the red line marks the best-fit curvature obtained by fitting the peak region.}
\label{fig:1100_max}
\end{figure}

\begin{figure}[!ht]
\center
 \includegraphics[width=18 cm, angle=0]{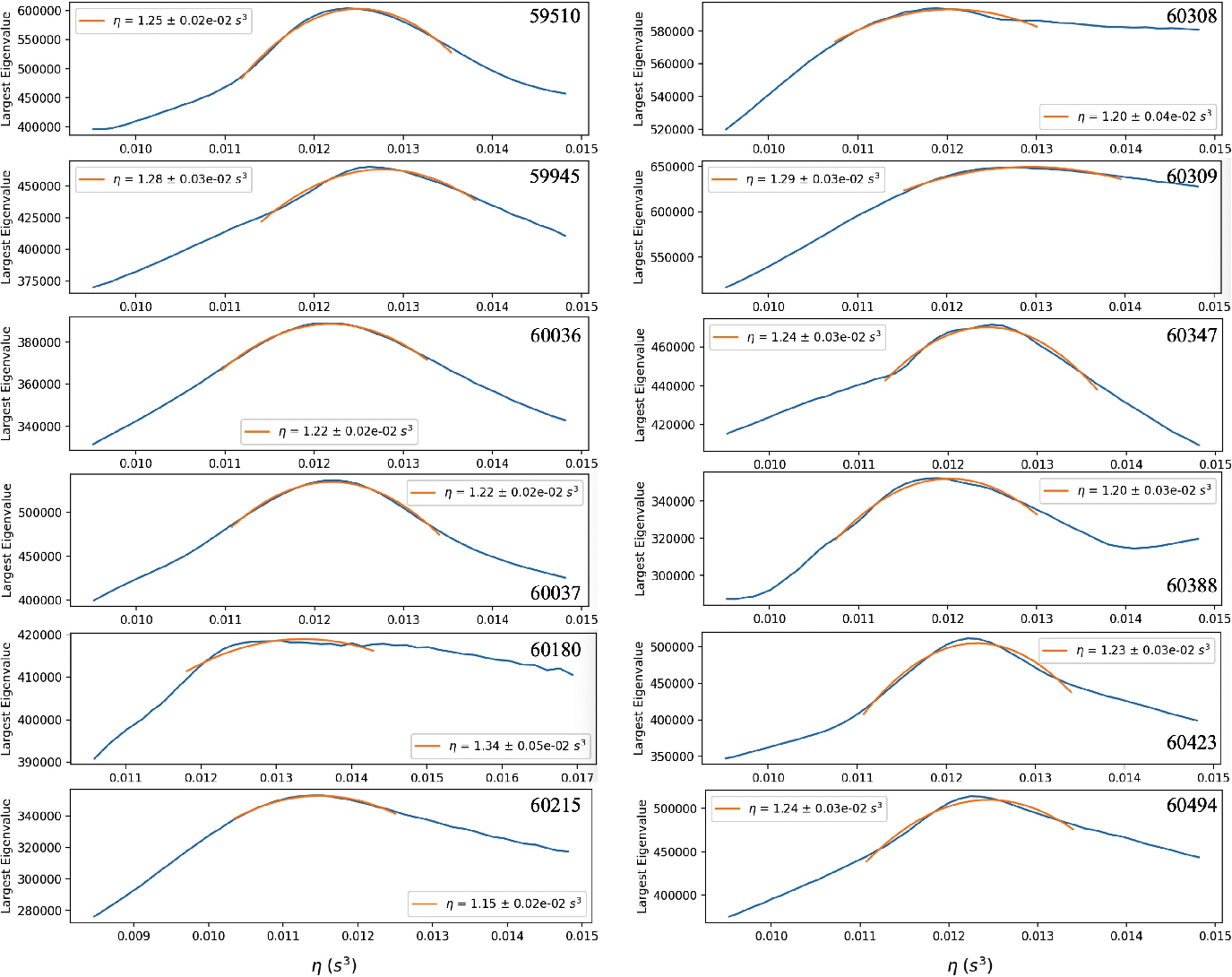}
    \caption{Same as the above figure, but showing the values of the largest eigenvalue from FAST observations centered at 1400~MHz.}
\label{fig:1400_max}
\end{figure}

\begin{figure}[!ht]
\center
 \includegraphics[width=8 cm, angle=0]{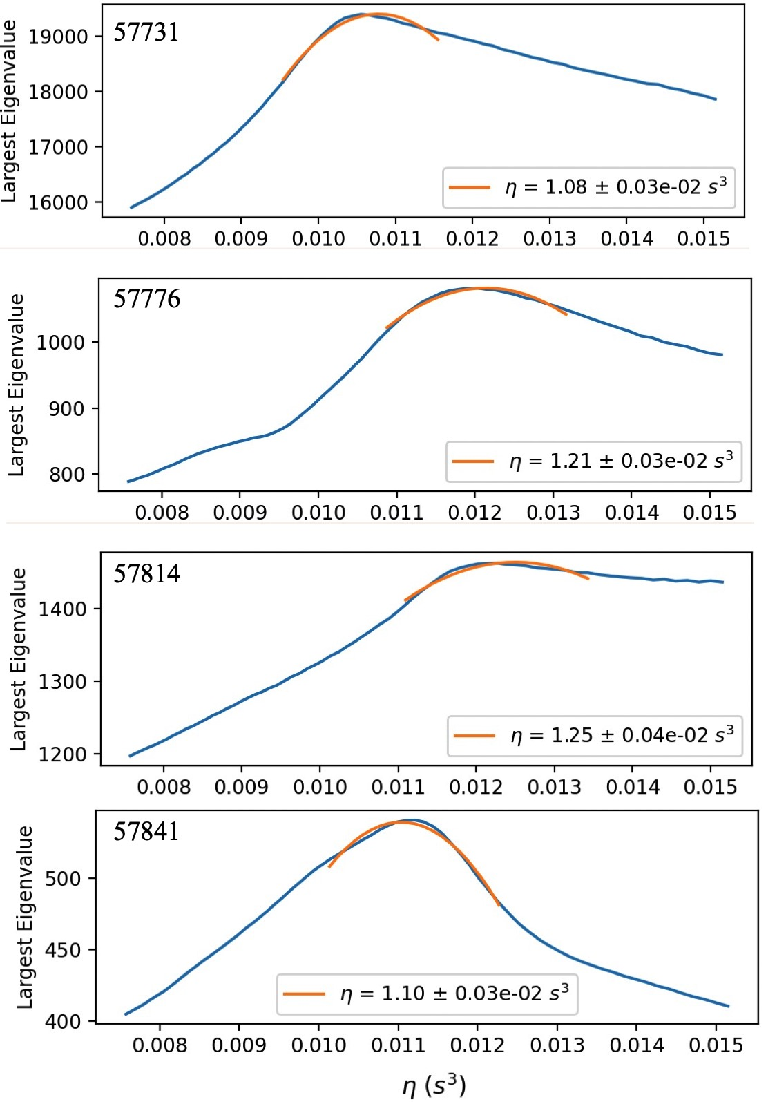}
    \caption{Same as the above figure, but showing the values of the largest eigenvalue from 4 Parkes observations centered at 1369~MHz.}
\label{fig:1369_max}
\end{figure}

Following \cite{swm+21} and \cite{bbk+22}, we obtained the distribution of the largest eigenvalue of the $\theta$–$\theta$ matrix for different arc curvatures across all observations from FAST and Parkes. As shown in Figures~\ref{fig:1100_max}, \ref{fig:1400_max}, and \ref{fig:1369_max}, the blue lines represent the distributions of eigenvalues evaluated at different arc curvatures. To determine the arc curvatures, we fitted a parabola to the peak region and obtained the best-fit curvature, indicated by the red lines.

\end{appendix}

\bibliographystyle{aasjournal}
\end{document}